\documentclass[aps,twocolumn]{revtex4}
\usepackage{epsfig}
\begin{document}
\def\I.#1{\it #1}
\def\B.#1{{\bf #1}}
\def\C.#1{{\cal  #1}}
\title{Transition in the Fractal Properties from Diffusion Limited Aggregation to 
Laplacian Growth via their Generalization}
   \author {H. George E. Hentschel$^*$, Anders Levermann and Itamar Procaccia}
   \affiliation{Department of~~Chemical Physics, The Weizmann Institute of
   Science, Rehovot 76100, Israel\\
$^*$ permanent address: Dept. of Physics, Emory University, Atlanta Ga.}
%
%
\begin{abstract}
We study the fractal and multifractal properties (i.e. the generalized dimensions 
of the harmonic measure) of a 2-parameter family of growth patterns that result from
a growth model that interpolates between Diffusion Limited Aggregation (DLA)
and Laplacian Growth Patterns in 2-dimensions. The two parameters are $\beta$ which determines
the size of particles accreted to the interface, and $\C.C$ which measures
the degree of coverage of the interface by each layer accreted to the growth pattern at
every growth step. DLA and Laplacian Growth are obtained at $\beta=0,~\C.C=0$
and $\beta=2,~\C.C=1$, respectively. The main purpose of this paper is to
show that there exists a line in the $\beta-\C.C$ phase diagram that separates
fractal ($D<2$) from non-fractal (D=2) growth patterns. Moreover, Laplacian
Growth is argued to lie in the non-fractal part of the phase diagram.
Some of our arguments are not rigorous, but together with the numerics
they indicate this result rather strongly. We first consider the family
of  models obtained for $\beta=0,~\C.C>0$, and derive for them a scaling
relation $D=2D_3$. We then propose that this family has growth
patterns for which $D=2$ for some $\C.C>C_{\rm cr}$, where $C_{\rm cr}$ may
be zero. Next we consider
the whole $\beta-\C.C$ phase diagram and define a line that separates
2-dimensional growth patterns from fractal patterns with $D<2$. We explain
that Laplacian Growth lies in the region belonging to 2-dimensional
growth patterns, motivating the main conjecture of this paper, i.e.
that Laplacian Growth patterns are 2-dimensional. The meaning of this
result is that the branches of Laplacian Growth patterns have finite 
(and growing) area on scales much larger than any ultra-violet cut-off length. 
\end{abstract}
   \maketitle
\section{Introduction}
In recent work \cite{00BDLP,01BDP} we have introduced a model of fractal growth processes
that interpolates between Diffusion Limited Aggregation (DLA) \cite{81WS} and
Laplacian Growth Patterns  \cite{58ST,84SB}, and employed this model to show that these
processes are not in the same universality classes. The aim of this
paper is to study the fractal properties of the resulting clusters.
In particular we will be led to conjecture that Laplacian Growth is
asymptotically of dimension 2, and in this sense is not a fractal at all.
This is in contradistinction to DLA for which the dimension had been
computed to be 1.713...\cite{00DLP}. 

Laplacian Growth Patterns are obtained when the boundary $\Gamma$ of
a 2-dimensional domain is grown
at a rate proportional to the gradient of a Laplacian field $P$.
Outside the domain
$\nabla^2 P=0$, and each point of $\Gamma$ is advanced at a rate
proportional to $\B.\nabla P$ \cite{58ST,84SB}. In Diffusion Limited
Aggregation (DLA) \cite{81WS} a 2-dimensional cluster is grown 
by releasing fixed size random walkers from infinity,
allowing them to walk around until they hit any particle belonging to
the cluster. Since the particles are released one by one
and may take arbitrarily long time to hit the cluster, the probability
field is quasi-stationary and in the
complement of the cluster we have again $\nabla^2 P=0$.
The boundary condition at infinity is the same for the
two problems; in radial geometry as $r\to \infty$ the flux
is $\B.\nabla P={\rm const}\times\hat r/r$. Since the probability
for a random walker to hit the boundary is again proportional to 
$|\B.\nabla P|$, one could think that in the asymptotic limit when the
size of the particle is much smaller than the radius of the cluster,
repeated growth events lead to a growth process which is
similar to Laplacian Growth. Of course,
the ultraviolet regularizations in the two processes were taken
different; in studying
Laplacian Growth one usually  
solves the problem with the boundary condition $P=\sigma\kappa$
where $\sigma$ is the surface tension and $\kappa$ the local
curvature of $\Gamma$ \cite{86BKT}. Without this 
(or some other) ultraviolet regularization
Laplacian Growth reaches a singularity (cusps) in finite time \cite{84SB}. In
DLA the ultraviolet regularization is provided by the finite size
of the random walkers. However, many researchers believed
\cite{84Pat} that this difference, which for very large
clusters controls only the smallest scales of the fractal patterns,
were not relevant, expecting the two models to lead to the clusters
with the same asymptotic dimensions. While we argued recently that 
the difference in ultraviolet regularization is indeed not crucial \cite{01BDP},
the two problems are nevertheless in two different universality
classes. To establish this we have
constructed a family of growth processes that includes
DLA and a discrete version of Laplacian Growth as extreme members, using the
same ultraviolet regularization (and see Sect. 2 for a further discussion
of the regularization).   We thus exposed the
essential difference between DLA and Laplacian Growth. DLA is grown serially,
with the field being updated after each particle growth. On the other
hand all boundary points of a Laplacian pattern are advanced in parallel
at once (proportional to $\B.\nabla P$). We showed that this
difference is fundamental to the asymptotic dimension, putting the
two problems in different universality classes \cite{00BDLP}. Here we wish to go
further and suggest that Laplacian Growth patterns are 2-dimensional.

In Sect.2 we review 
briefly the two parameter model that had been introduced to establish
these results. We discuss there the two parameters $\beta$ and $\C.C$ that are
used to interpolate between DLA and Laplacian Growth. In Sect.3 we 
analyze the generalized dimensions
$D_q$ and relate them to the scaling of moments of objects which are natural
to the theory. In Sect. 4 we discuss first a family of growth models
which is a 1-parameter generalization of DLA, ($\beta=0$, $0\le \C.C\le 1$,
and show that the fractality of DLA is lost for some $\C.C>\C.C_{\rm cr}$
in favor of 2-dimensional growth patterns. It is not impossible that
$\C.C_{\rm cr}=0$. For growth patterns
in this family we derive a scaling relation $D=2D_3$. Under some
plausible assumptions we propose that for $\C.C>\C.C_{cr}$ there
exists another scaling relation, i.e. $D=1+D_2$, which implies
immediatly that $D=2$. Secondly we discuss
the 1-parameter family of models that generalizes Laplacian
Growth ($\beta=2$, $0\le \C.C\le 1$ and show that the above 
relation is not obtained here, leading to the existence of
fractal patterns also for high values of $\C.C$.  Finally in Sect. 5 we reach the main
conjecture of this paper, i.e. that Laplacian Growth patterns are
2-dimensional.  In Sect. 6 we offer a discussion and some open questions
that are left for future research.
\section{Iterated Conformal Maps for Parallel Growth Processes}

The method of iterated conformal maps for DLA was introduced in \cite{98HL}. 
In \cite{00BDLP,01BDP} we have presented a generalization
to parallel growth processes.
We were interested in
$\Phi^{(n)}(w)$ which conformally maps the exterior of the unit
circle $e^{i\theta}$ in the
mathematical $w$--plane onto the complement of the (simply-connected)
cluster of $n$ particles in the physical $z$--plane.
The unit circle is mapped onto the boundary of the cluster. 
In what follows we use the fact that the gradient of the Laplacian field
$\B.\nabla P(z(s))$ is 
\begin{equation}
|\B.\nabla P(z(s))| =\frac{1}{|{\Phi^{(n)}}^\prime(e^{i\theta})|}\ ,\quad
z(s)=\Phi^{(n)}(e^{i\theta})
\ .
\label{gradient}
\end{equation}
Here $s$ is an arc-length parametrization
of the boundary.
The map $\Phi^{(n)}(w)$ is constructed recursively. Suppose that
we have already $\Phi^{(n)}(w)$ which maps to the exterior of a cluster
of $n$ particles in the physical
plane and we want to 
find the map $\Phi^{(n+p)}(w)$ after $p$ additional particles were added to its 
boundary {\em at once}, each
proportional in size to the local value of $|\B.\nabla P|^{\beta/2}$.
To grow {\em one} such particle we employ the elementary map 
$\phi_{\lambda,\theta}$ which transforms the unit
circle to a circle with a semi-spherical ``bump" of linear size $\sqrt{\lambda}$
around the point $w=e^{i\theta}$:
\begin{eqnarray}
   &&\phi_{\lambda,0}(w) = \sqrt{w} \left\{ \frac{(1+
   \lambda)}{2w}(1+w)\right. \nonumber\\
   &&\left.\times \left [ 1+w+w \left( 1+\frac{1}{w^2} -\frac{2}{w}
\frac{1-\lambda} {1+ \lambda} \right) ^{1/2} \right] -1 \right \} ^{1/2} \\
   &&\phi_{\lambda,\theta} (w) = e^{i \theta} \phi_{\lambda,0}(e^{-i
   \theta}
   w) \,,
   \label{eq-f}
\end{eqnarray}
If we update the field after the addition of this single particle,
then
\begin{equation}
   \Phi^{(n+1)}(w) = \Phi^{(n)}(\phi_{\lambda_{n+1},\theta_{n+1}}(w)) \ ,
   \label{recurs}
\end{equation}
where $ \Phi^{(n)}(e^{i \theta_{n+1}})$ is the point on which the 
$(n+1)$-th particle is grown and 
$\sqrt{\lambda_n}$ is the size of the grown particle 
divided by the Jacobian of the map, ${\Phi^{'(n)}}(e^{i \theta_{n+1}})$,
at that point.
  
The map $\Phi^{(n+1)}(w)$ adds on a
new semi-circular bump to the image of the unit circle under
$\Phi^{(n)}(w)$. The bumps in the $z$-plane simulate the accreted 
particles in
the physical space formulation of the growth process. For the height
of the bump to be proportional to $|\B.\nabla P(z(s))|^{\beta/2}$ we need to
choose its area proportional to
$|{\Phi^{(n-1)}}' (e^{i \theta_n})|^{-\beta}$ (see Eq. (\ref{gradient})),
or
\begin{equation}
\lambda_{n} = \frac{\lambda_0}{|{\Phi^{(n-1)}}' (e^{i \theta_n})|^{\beta+2}}
\ . \label{lambdanew}
\end{equation}
With $\beta =0$ these rules produce a DLA cluster, for which the
particles are of constant area. With $\beta=2$ we
grow bumps in the physical space whose linear scale is proportional the gradient of the field
$|\B.\nabla P(z(s))|$, as is appropriate for Laplacian Growth.
Next, to grow $p$ (non-overlapping) particles in parallel, we accrete them without
updating the conformal map. In other words, to add a new layer of $p$
particles when the cluster contains $m$ particles, we need to choose $p$
angles on the unit circle $\{\tilde\theta_{m+k}\}_{k=1}^p$. At these
angles we grow bumps which in the physical space have the wanted
linear scale (ranging from constant to proportional to the gradient
of the field):
\begin{equation}
   \lambda_{m+k} = \frac{\lambda_0}{|{\Phi^{(m)}}'
   (e^{i \tilde\theta_{m+k}})|^{\beta+2}} \ , \quad k=1,2\dots, p \ .
\label{layer}
\end{equation} 
After the $p$ particles were added, the conformal map and thus the field should
be updated. In updating, we will use $p$ compositions of the elementary
map $\phi_{\lambda,\theta} (w)$. Of course, every composition 
effects a reparametrization of the unit circle, which
has to be taken into account. To do this, we
define a series
$\{\theta_{m+k}\}^p_{k=1}$ according to
\begin{equation}
   \Phi^{(m)}(e^{i\tilde\theta_{m+k}})\equiv
   \Phi^{(m+k-1)}(e^{i\theta_{m+k}})\ . \label{tilde}
\end{equation}
Next we define the conformal map used in the next layer growth
according to
\begin{equation}
   \Phi^{(m+p)}(\omega)\equiv
\Phi^{(m)}\circ\phi_{\theta_{m+1},\lambda_{m+1}}\circ
   \dots\circ
   \phi_{\theta_{m+p},\lambda_{m+p}}(\omega) \ . \label{compose}
\end{equation}
In this way we achieve the growth at the images under $\Phi^{(m)}$ of the
points $\{\tilde\theta_{m+k}\}_{k=1}^p$. To compute the $\theta$ series from
a given $\tilde \theta$ series we use Eq.(\ref{compose}) to rewrite
Eq.(\ref{tilde}) in the form
\begin{equation}
e^{i\theta_{m+k}} =\phi^{-1}_{\theta_{m+k-1},\lambda_{m+k-1}}\circ
\dots \circ \phi^{-1}_{\theta_{m+1},\lambda_{m+1}}(e^{i\tilde
  \theta_{m+k}})
\label{connection}
\end{equation}
The inverse map $\phi^{-1}_{\theta,\lambda}$ is given by 
$\phi^{-1}_{\theta,\lambda}(\omega)=
e^{i\theta}\phi^{-1}_{0,\lambda}(e^{-i\theta}\omega)$ with
\begin{equation}
\phi^{-1}_{0,\lambda} = \frac
{\lambda \omega^2  \pm \sqrt{\lambda^2 \omega^4  - 
\omega^2 [1-(1+\lambda)\omega^2][\omega^2-(1+\lambda)]}}
{1-(1+\lambda)\omega^2} \ ,
\label{inverse-map}
\end{equation}
where the positive root is taken for Re $\omega > 0$ and the negative root 
for Re $\omega < 0$. 

Evidently, Laplacian Growth 
calls for choosing the series
$\{\tilde\theta_{m+k}\}_{k=1}^p$ such as to have full coverage
of the unit circle (implying the same for the boundary $\Gamma$).
On the other hand DLA calls for growing a single particle 
before updating the field. Since it was shown \cite{99DHOPSS} that in DLA
growth $\lambda_n$ decreases on the average when $n$ increases, in 
the limit of large clusters DLA is consistent with vanishingly small coverage of the
unit circle. To interpolate between these two cases we
introduce a parameter that serves to distinguish one growth model from
the other, giving us a 2-parameter control (the other parameter is $\beta$). This 
parameter is the {\em degree of coverage}. Since the area covered by the 
pre-image of the $n$-th particle on the unit circle is approximately
$2 \sqrt{\lambda_n}$, we introduce the parameter
\begin{equation}
   {\cal C}=\frac{1}{\pi}\sum_{k=1}^p \sqrt{\lambda_{m+k}} \ . \label{defC}
\end{equation}
(In \cite{01BDP} we showed how to measure the coverage exactly).
Since this is the fraction of the unit circle which is covered in
each layer, the limit of Laplacian Growth is obtained with ${\cal C}=1$.
DLA is asymptotically consistent with ${\cal C}=0$. Of course,
the two models differ also in the size of the growing bumps, with
DLA having fixed size particles, ($\beta=0$ in Eq.(\ref{lambdanew})), 
and Laplacian Growth having particles proportional to $\B.\nabla P$ 
($\beta=2$ in Eq.(\ref{layer})). 
Together with ${\cal C}$ we have a two parameter control on the
parallel growth dynamics, with DLA and Laplacian Growth 
occupying two corners of the $\beta, {\cal C}$ plane, at the points
(0,0) and (2,1) respectively.

Obviously, the partially serial growth within the layer introduces
an additional freedom which is the {\em order} of placement of 
the bumps on the unit circle. In \cite{00BDLP,01BDP}
we have shown that the order is in fact immaterial as far 
as the asymptotic fractal properties
of the clusters are concerned. Accordingly, we will take
random choices of $\tilde\theta_{m+k}$
with a rule of skipping overlaps.

We should note that in our approach the regularization of putative
singularities is not achieved with surface tension, but by having a minimal
size bump, similarly to the regularization of DLA. Our rules of growth
with $\beta=2$ and $\lambda_0$ chosen once and for all, guarantee that every 
layer of growth has exactly the
same area. This in the continuous time Laplacian Growth model translates to 
a particular choice of the time step $dt$. Clearly, one has freedom in choosing
$dt$, or of the size $\lambda_0$ in each layer, as long as this does
not affect the nature of the growth. In particular we can have $\lambda_0$ chosen 
such that the maximal
physical bump is of constant area.  Once $\lambda_0$ is chosen, 
the sharpest feature that can be achieved is
a bump of size $\lambda_0$, and the worst possible ``singularity" is a line
of such bumps, exactly as in DLA. Thus the putative cusp singularity of
Laplacian Growth is avoided in a manner that is identical for all the
growth models in our 2-parameter family. 
 
The conformal map
$\Phi^{(n)}(\omega)$ admits a Laurent expansion
\begin{equation}
\label{Laurent}
   \Phi^{(n)}(\omega) = F_1^{(n)} \omega +F_0^{(n)}
   +\frac{F_{-1}^{(n)}}{\omega}+~\cdots \ .
\end{equation}
The coefficient of the linear term is the Laplace radius, and was
shown to scale like
\begin{equation}
\label{scaleF1}
   F_1^{(n)}\sim S^{1/D} \ ,
\end{equation}
where $S$ is the area of the cluster,
\begin{equation}
\label{defS}
S= \sum_{j=1}^n \lambda_j~ |{\Phi^{'(j-1)}} (e^{i\theta_j})|^2 \ .
\end{equation}
Note that for $\beta=0$ this and equation (\ref{lambdanew}) imply
that $S=n\lambda_0 $. Indeed for $\beta=0$ this estimate had
been carefully analyzed and substantiated (up to a factor) in \cite{2001SL}.
On the other hand $F_1^{(n)}$ is given analytically by
\begin{equation}
   F_1^{(n)} = \prod_{k=1}^n \sqrt{(1+\lambda_k)} \ ,
\end{equation}
and therefore can be determined very accurately. 

The conclusion from the 
calculations presented in \cite{00BDLP,01BDP} is that for $\C.C>0$ 
the fractal dimension of the growth patterns
depends continuously on the parameters, growing monotonically
upon decreasing $\beta$ or increasing ${\cal C}$. It is quite obvious why
increasing ${\cal C}$ should increase the dimension. By forbidding
particles to overlap we simply force them
into the fjords, not allowing them to hit the tips only (as is highly probable).
Also decreasing $\beta$ increases the dimension, since we grow larger 
particles into the fjords, whereas
increasing $\beta$ reduces the size of particles added to fjords and increases the size of particles
that accrete onto tips. In particular
we argued that DLA and our discretized Laplacian Growth cannot have the same
dimensions, putting them in different universality classes.
In the rest of this paper we make these observations more 
quantitative and precise.
\section{Multifractal Properties}
\subsection{Generalized Dimensions} 
The fractal dimension in the $\beta-{\cal C}$ family of models, $D(\beta ,C)$,
 is introduced as the
exponent relating the area of the cluster $S_n$ to its linear scale (which
is measured by the (dimensionless) Laplace radius $F_1^{(n)}$):
 \begin{equation}
 \label{fractal}
 S_n \sim (F_1^{(n)})^{D(\beta ,C)} \tilde \lambda_0 \ .
 \end{equation}
 In this equation $\tilde \lambda_0 \equiv \lambda_0^{2/(2+\beta)}$. 
The multifractal exponents
\cite{83HP} are defined in analogy to those for DLA in terms of the moments of the 
(dimensionless) electric field $E(s)$ on
the boundary of the cluster \cite{86HMP},
 \begin{eqnarray}
 \label{multifractal}
 \langle E^{(q-1)}\rangle &\sim& (F_1^{(n)})^{-(q-1)D_q(\beta ,C)}
 \nonumber\\ &\sim& (S_n/\tilde \lambda_0)^{-(q-1)D_q(\beta ,C)/D(\beta ,C)},
 \end{eqnarray}
 where $\langle \cdots \rangle$ represents the harmonic average for the 
$(\beta , C)$ clusters in question. Note that these exponents are for a 
fixed size partition with boxes of length $\sqrt{\tilde\lambda_0}$, with asymptotics for an
infinitely large cluster. A supremum over arbitrary partitions may
lead to different exponents, cf. \cite{86HJPKS}.

 This result translates immediately \cite{99DHOPSS} to the multifractal fluctuations of 
the bump areas $\lambda_n$ added
in the mathematical plane. As $\lambda_n^q 
 \sim  E_n^{(2+\beta )q}$, where $E_n$ is the field computed at $z(s)=\Phi^{(n)}(e^{i\theta_n})$.
We therefore write
 \begin{equation}
 \label{lambda}
 \langle \lambda_n^q \rangle 
 \sim (S_n/\tilde \lambda_0)^{-(2+\beta )q D_{(2+\beta )q+1}(\beta ,C)/D(\beta ,C)}.
 \end{equation}
 Specifically we can derive the following important moments
 \begin{eqnarray}
 \label{moments}
\langle \sqrt{\lambda_n} \rangle & \sim & (S_n/\tilde\lambda_0)^{-(1+\beta /2) D_{(2+\beta /2)}/D}
\nonumber \\ 
 \langle \lambda_n \rangle & \sim &(S_n/\tilde\lambda_0)^{-(2+\beta ) D_{3+\beta}/D}\\
 \langle \lambda_n^{\beta /(2+\beta) } \rangle & \sim & (S_n/
\tilde\lambda_0)^{-\beta D_{1+\beta}/D}\nonumber
\\
 \end{eqnarray}
 where naturally all the dimensions are functions of $(\beta ,C)$.
 We can also estimate the way in which the maximal bump areas scale
 \begin{equation}
 \label{maximum}
 \lambda_{n,max}\equiv \lim_{q\to \infty}  \langle \lambda_n^q \rangle^{1/q} \sim 
(S/\tilde\lambda_0)^{-(2+\beta ) D_{\infty}(\beta ,C)/D(\beta ,C)}.
 \end{equation}

 Consider now the addition of one layer of $p$ particles 
to the growing cluster. We can rewrite Eq.~(\ref{defC}) as 
 \begin{equation}
 \label{C2}
 C = (1/\pi ) p \overline{\sqrt{\lambda_{n}}},
 \end{equation}
where we have introduced the notation $\overline{\lambda_n^q}$ to represent 
the average over a layer of $p$ particles:
\begin{equation}
\overline{f(\lambda_n)}\equiv \frac{1}{p}\sum_{k=1}^p f(\lambda_{n+k})
\end{equation} For our considerations
below it is important to relate the layer averages $\overline{\lambda_n^q}$ 
to harmonic averages $\langle \lambda_n^q \rangle$. This relationship
may very well depend on the value of $\beta$. The two case that 
are of highest interest to us are $\beta=0$ and $\beta=2$, and
we will examine them separately.
\section{Scaling relations for the Fractal Dimension $D$}
\subsection{The case $\beta=0$ and ${\cal C}>0$}
\begin{figure}
\centering
\includegraphics[width=.33\textwidth]{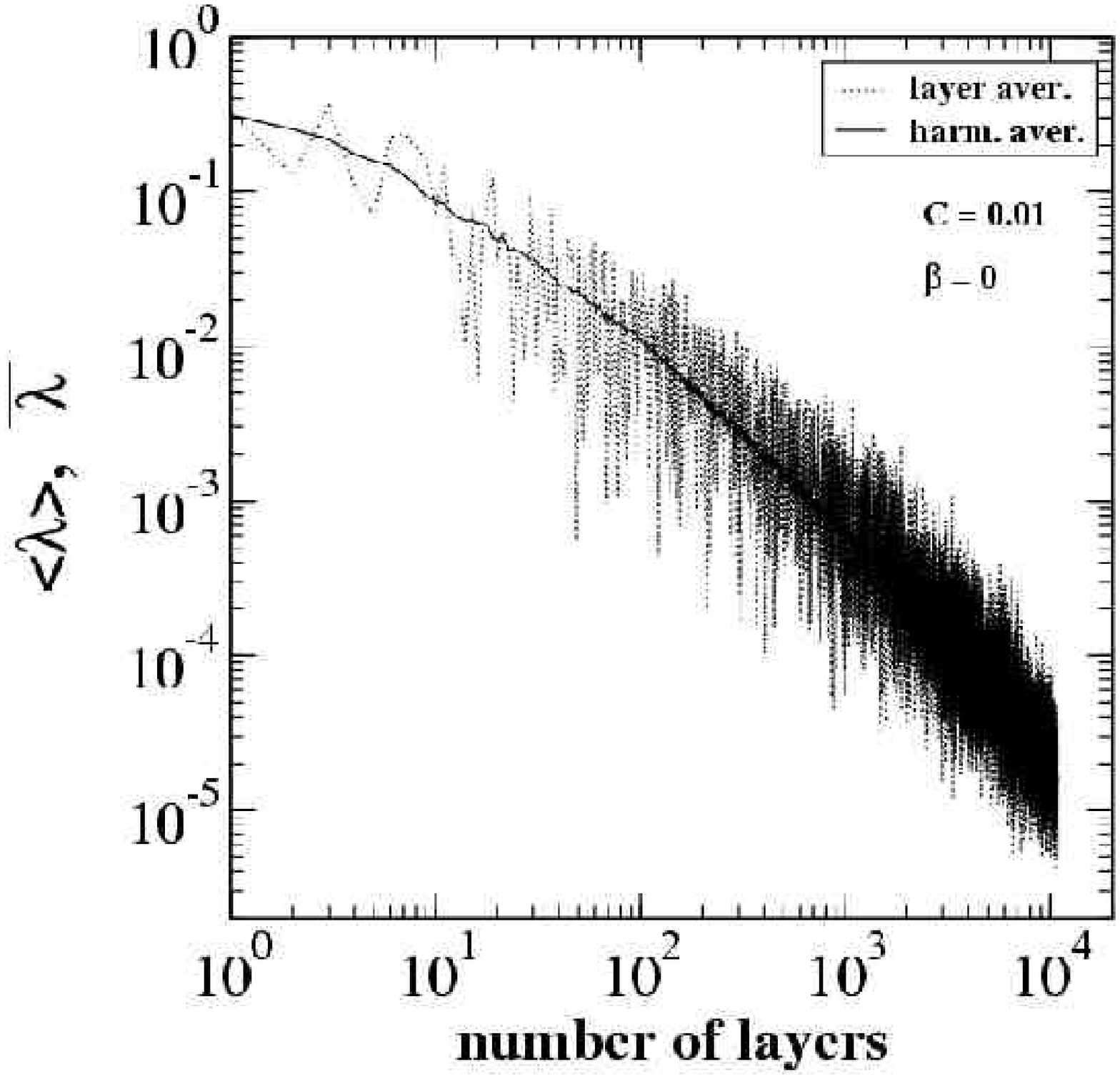}
\includegraphics[width=.33\textwidth]{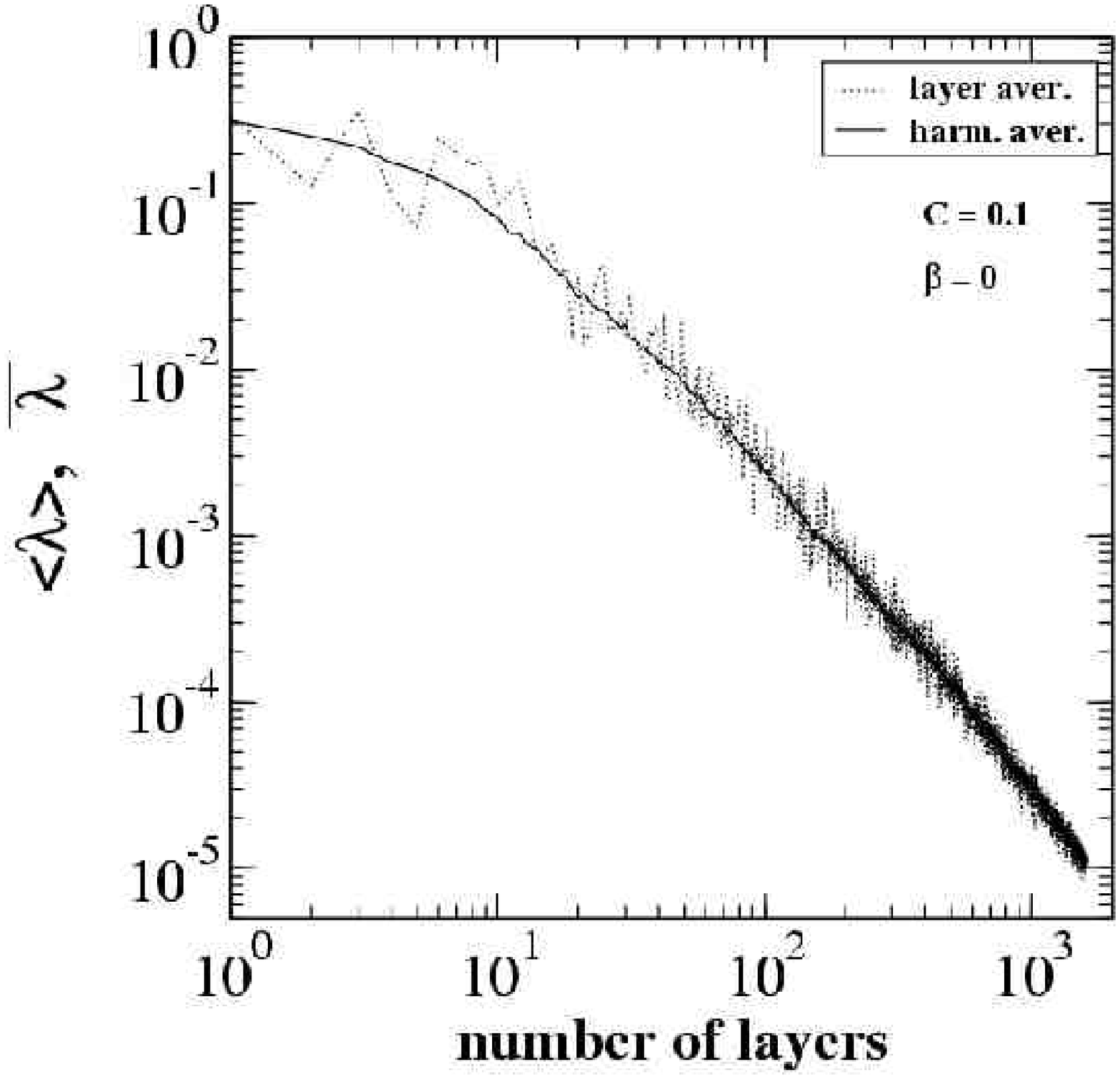}
\includegraphics[width=.33\textwidth]{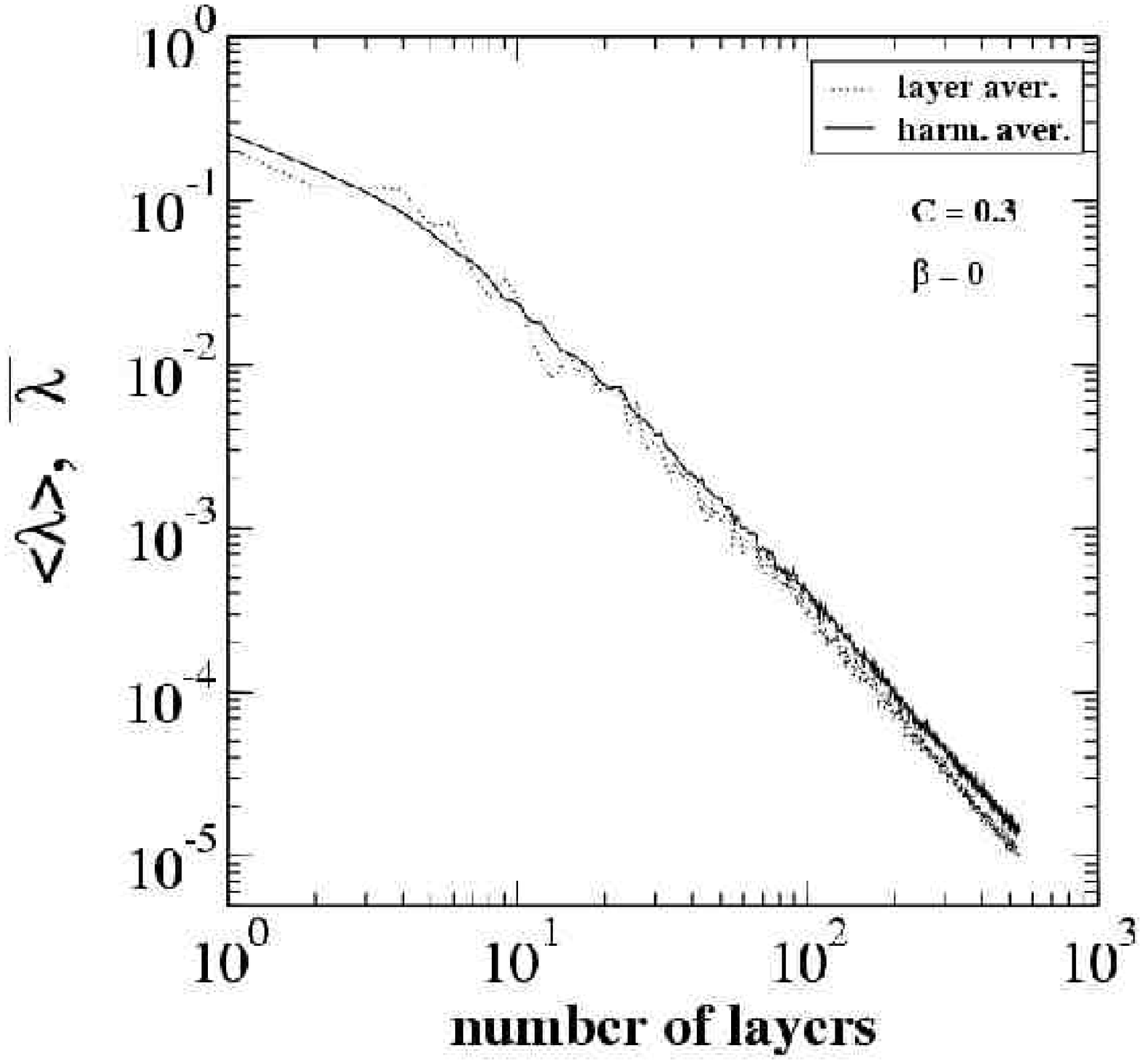}
\includegraphics[width=.33\textwidth]{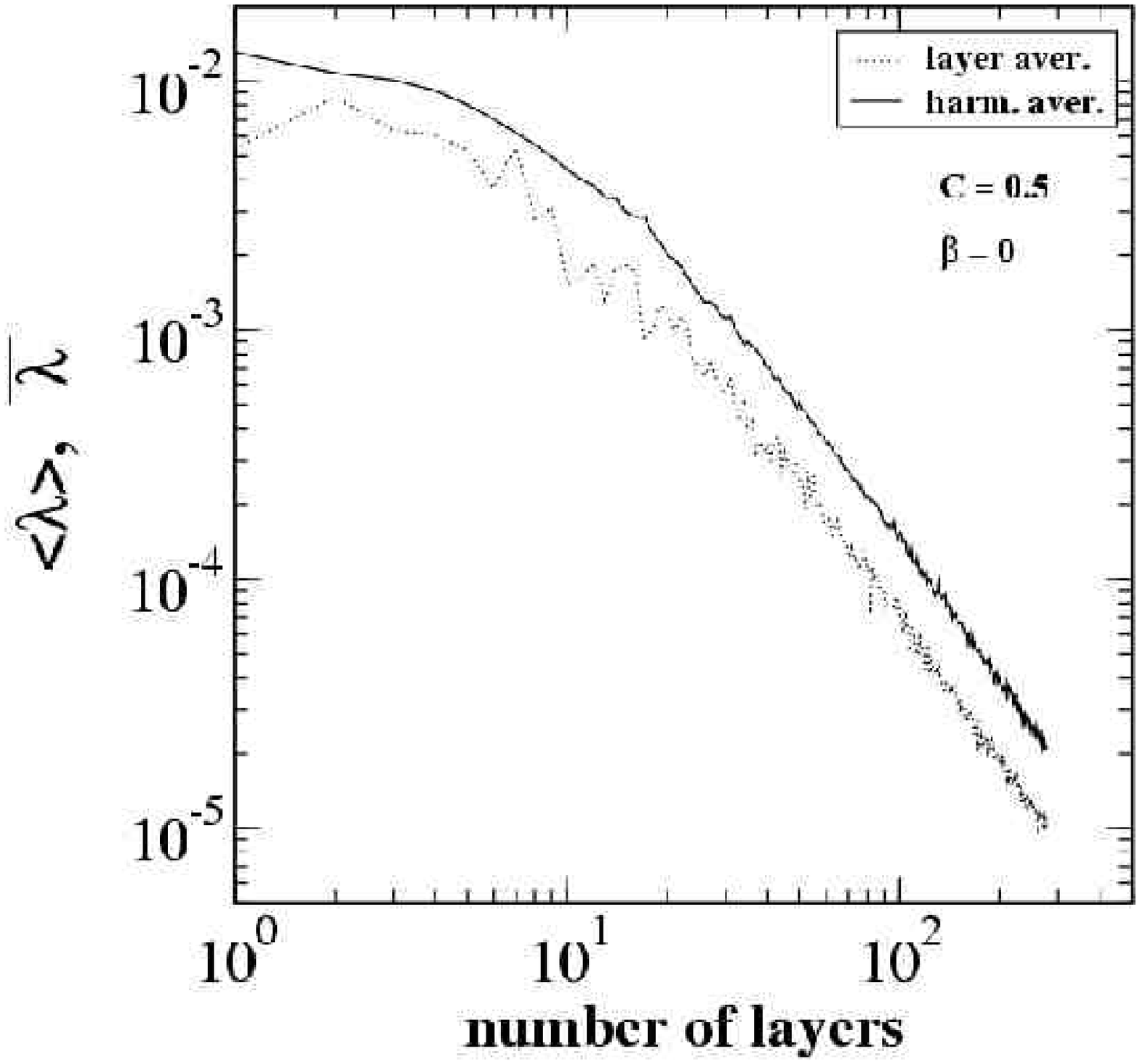}
\caption{Layer and harmonic averages of $\lambda_n$ as a function of the
number of layers, for $\beta=0$. Panels a-d: ${\cal C}=0.01,
0.1,0.3$ and 0.5 respectively.}
\label{averages01}
\end{figure}

\begin{figure}
\centering
\includegraphics[width=.33\textwidth]{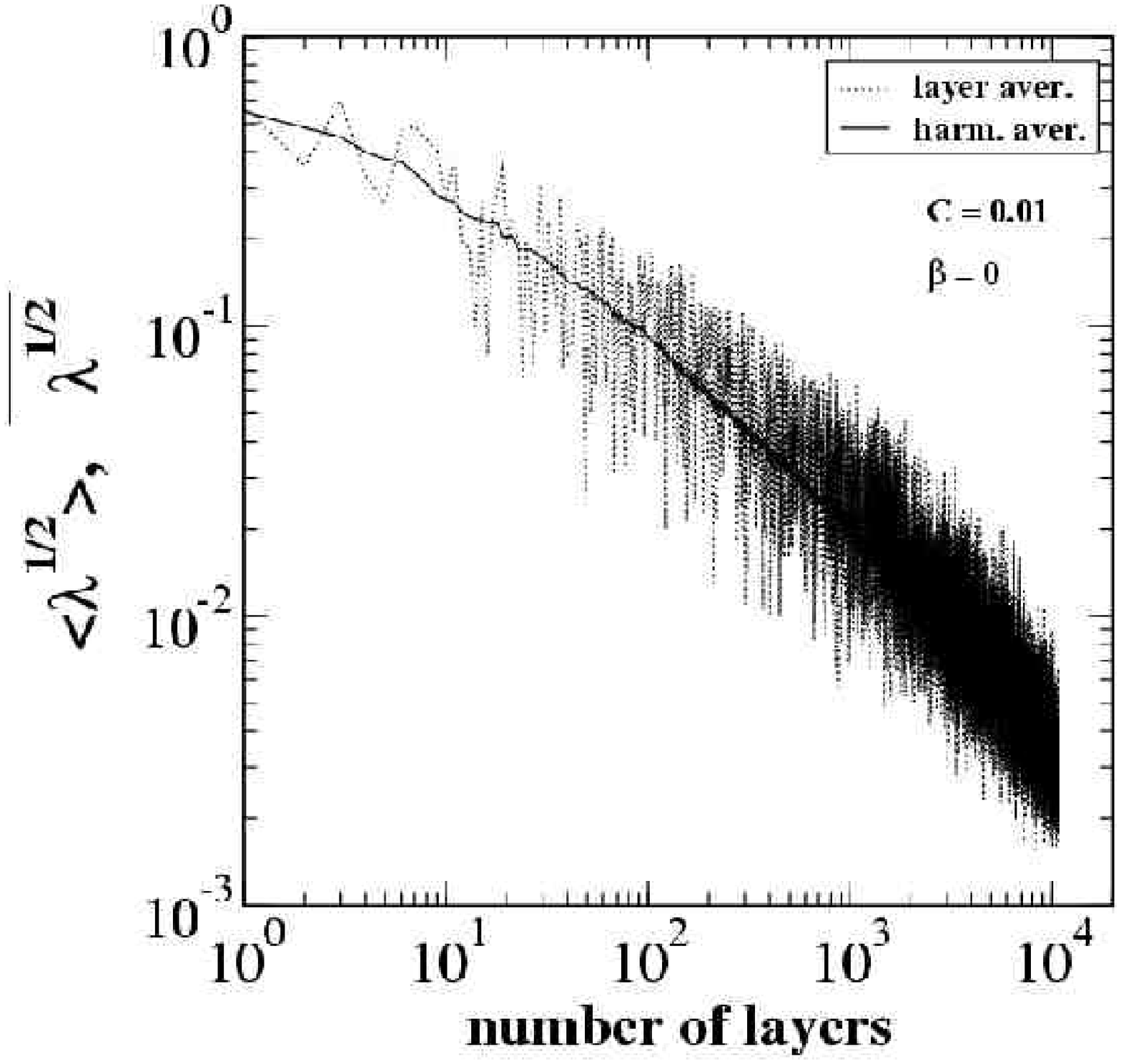}
\includegraphics[width=.33\textwidth]{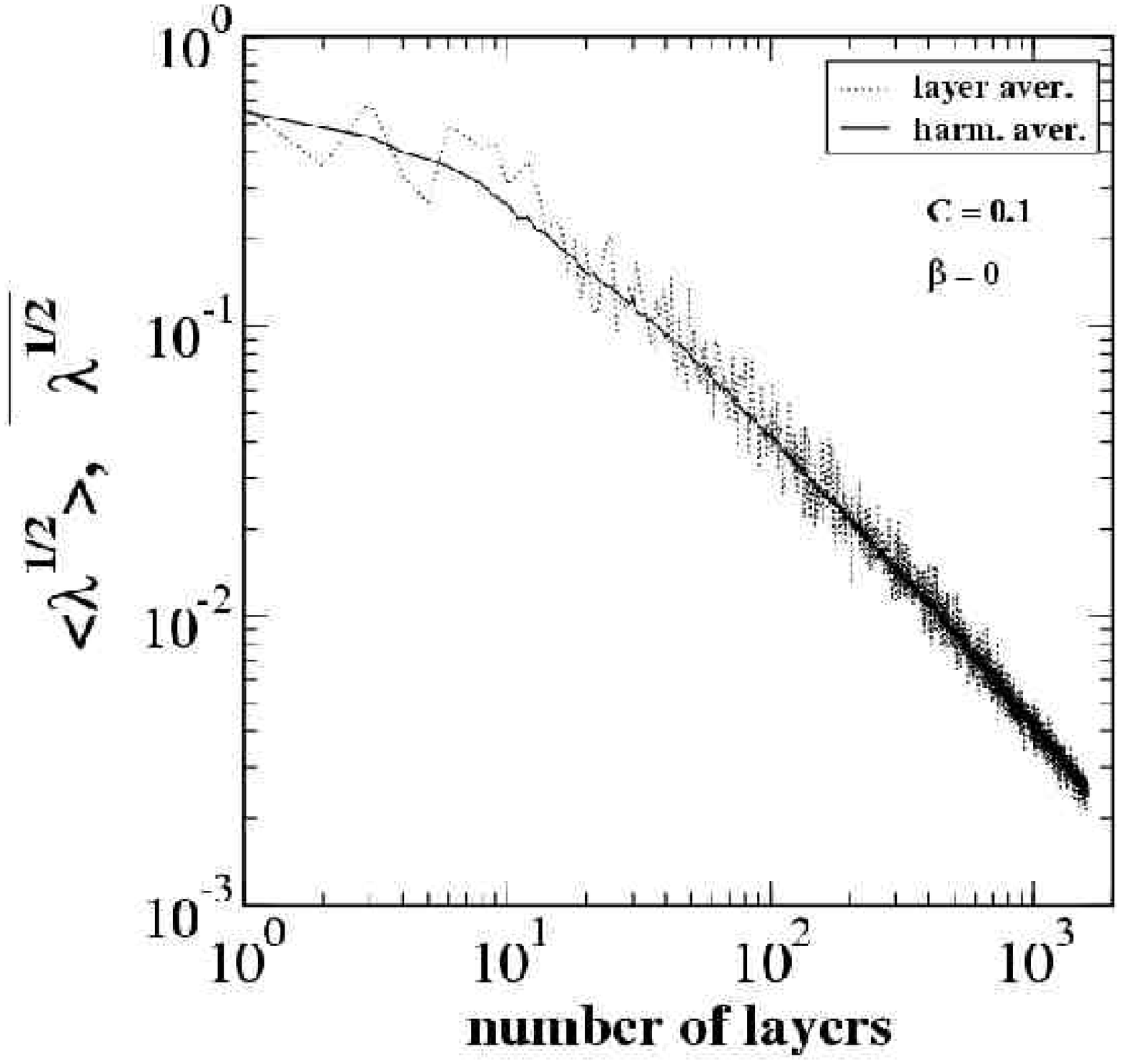}
\includegraphics[width=.33\textwidth]{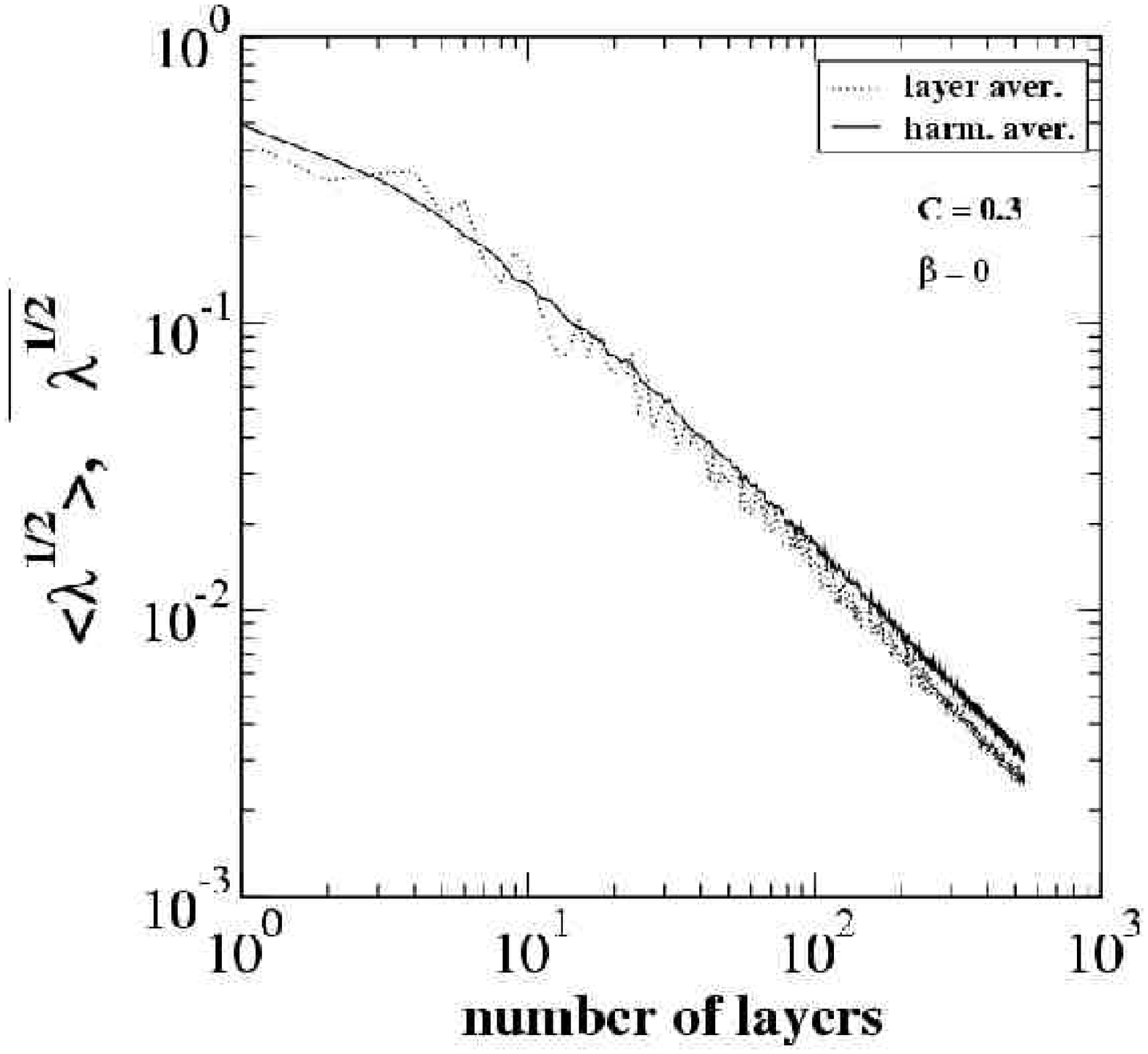}
\includegraphics[width=.33\textwidth]{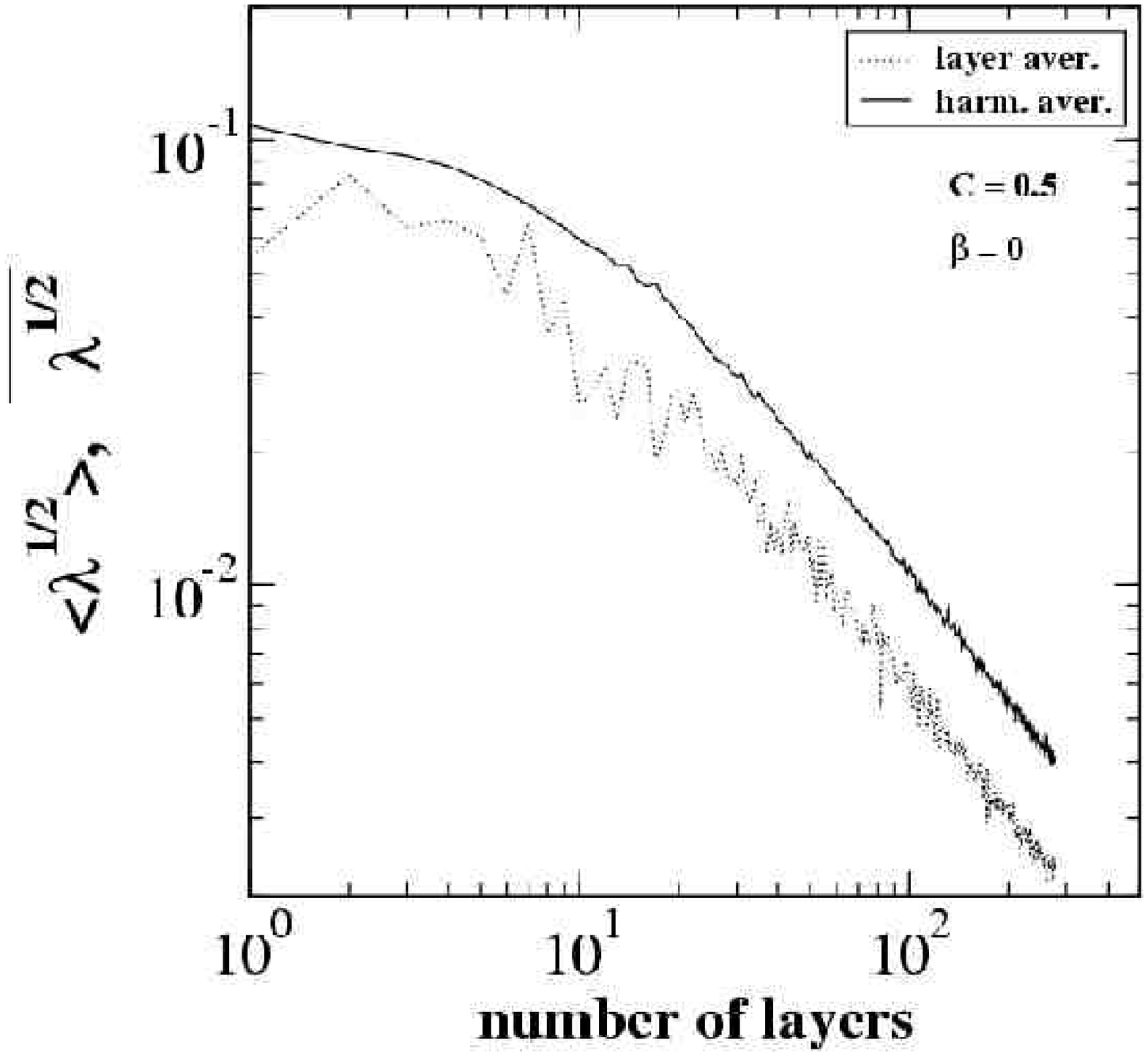}
\caption{Layer and harmonic averages of $\lambda^{0.5}_n$ as a function of the
number of layers, for $\beta=0$. Panels a-d: ${\cal C}=0.01,
0.1,0.3$ and 0.5 respectively.}
\label{averages005}
\end{figure}

We examine the relationship between layer and harmonic averages
numerically. In Fig. \ref{averages01} we show the two averages 
vs. the number of layers for the case $q=1$, $\beta=0$ and four values
of ${\cal C}$. In Fig. \ref{averages005} we show the same
for the case $q=0.5$, $\beta=0$ and the same four values
of ${\cal C}$. 
Examining the results it appears that for the
higher values of $\C.C$ we can assume that 
in the scaling sense
 \begin{equation}
 \label{averages}
 \overline{\lambda_n^q} \sim  \langle \lambda_n^q \rangle\ , \quad \beta=0 \ .
 \end{equation}
Note that for smaller values of ${\cal C}$ the evidence
is not as clear cut as for higher values. The number of points $p$ 
in each layer is relatively
small and the layer average is highly fluctuating. Nevertheless,
even for the case ${\cal C}=0.01$, if we perform a running average
on the layer average data, we converge very well onto the harmonic
average. We therefore propose to proceed with the conjecture that
Eq. (\ref{averages}) is correct for all the values of ${\cal C}$
and $\beta=0$, and
investigate the implications of this scaling relation for the cases
for which it is correct.  An immediate consequence
 of Eqs.~(\ref{C2}) and (\ref{averages}) is that
 \begin{equation}
 \label{layernumber}
 p \sim {\cal C}/\langle \sqrt{\lambda_n}\rangle 
\sim {\cal C}(S/\tilde\lambda_0)^{ D_2/D}
\end{equation}
We note that this means that $p\to \infty$ asymptotically
for every value of ${\cal C}$, while $p/n\to 0$.

Next observe that by definition
\begin{equation}
F_1^{(n+p)}/F_1^{(n)}=\Pi_{k=1}^p(1+\lambda_{n+k})^a\approx 1+ap\overline{\lambda_n} \ .
\label{Fratio}
\end{equation}
In light of Eq. (\ref{fractal}) we write
\begin{equation}
\frac{S_{n+p}}{S_n}=\left(\frac{F_1^{(n+p)}}{F_1^{(n)}}\right)^D
\approx 1+aDp\overline{\lambda_n} \ . \label{Sratio}
\end{equation}
On the other hand we estimate
\begin{equation}
\frac{S_{n+p}}{S_n}\approx 1+\frac{p\tilde\lambda_0}{S_n} \ , \label{Sratio2}
\end{equation}
and comparing with (\ref{Sratio}) we find
\begin{equation}
S_n \approx \frac{\tilde\lambda_0}{aD\overline{\lambda_n}} \ . \label{Sapprox}
\end{equation}
If Eq.(\ref{averages}) is used, we find finally
\begin{equation}
S_n \approx \tilde\lambda_0 \left(\frac{S_n}{\tilde\lambda_0}\right)^{2D_3/D} \ . \label{Sfinal}
\end{equation}
from which we derive the well known ``electrostatic relation"
\begin{equation}
D=2D_3 \ . \label{electro}
\end{equation}
This result was known for ${\cal C}=0$ \cite{94Hal}, and is generalized here, under the
conjecture (\ref{averages}) to all values of ${\cal C}$.

Let us consider now the probability to hit at the point of maximal radius. 
We propose that for any finite
${\cal C}$ the probability for this event is finite. We stress that this
``point" is actually a region on the interface of size $\sqrt{\tilde\lambda_0}$ in every
layer. In particular, we expect that the growth process will hit the point of
maximal radius every finite number of layers, where this number is of the
order of $1/{\cal C}$.  We also know for sure that we have at most one hit per 
layer since particles cannot overlap in the dynamics.

Consider now the scaling of the size of the growth pattern which is measured by $F_1^{(n)}$. 
First we know that $F_1^{(n)}\sim (S/\tilde\lambda_0)^{1/D}$, and therefore
\begin{equation}
\label{scaling1}
dF_1^{(n)}/dS \sim (S/\tilde\lambda_0)^{1/D -1}/\tilde\lambda_0 \ . 
\end{equation}
On the other hand, we estimate the same object using the following
argument: the maximal radius $R^{(n)}$ increases by 
$\sqrt{\tilde\lambda_0}$ every time that it is hit. 
This occurs every $1/{\cal C}$
layers in which $p$ particles were added. Therefore
\begin{equation}
\label{scaling2}
\frac{dR^{(n)}}{dS} \sim \frac{\sqrt{\tilde\lambda_0}}{p {\tilde\lambda_0}/{\cal C}}
\end{equation}
Comparing Eqs. (\ref{scaling1})
and (\ref{scaling2}), using Eq. (\ref{layernumber}) we obtain the scaling relation
\begin{equation}
\label{D2}
D = 1 + D_2
\end{equation}
Using the inequalities between the generalized dimensions and Eq.
(\ref{electro}) we write
\begin{equation}
D-1= D_2 \ge D_3 = D/2 \quad {\rm for~ all}~ {\cal C}>\C.C_{\rm cr} \ , \label{Dineq}
\end{equation}
which is equivalent to 
\begin{equation}
D=2 \quad {\rm for~ all}~ {\cal C}>\C.C_{\rm cr} 
\end{equation} 
In other words, we conclude that along the line $\beta=0$
in the phase diagram $\beta-\C.C$, there exists a transition
to growth patterns of dimension 2. 

\begin{figure}
\centering
\includegraphics[width=.33\textwidth]{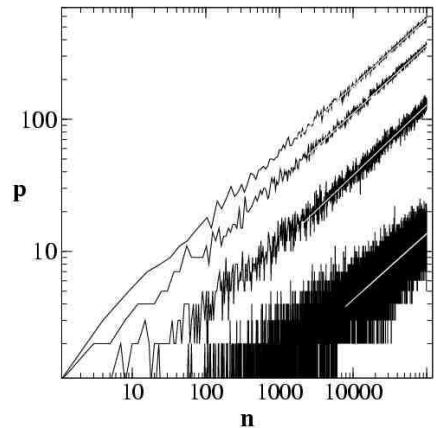}
\caption{The number $p$ of bumps in a layer vs. the number $n$ of
bumps in the growth pattern, in a log-log plot. From
top to bottom that is shown for $\C.C=0.5$, 0.3, 0.1 and 0.01 respectively.
In white lines we show the scaling laws $p\sim n^{1/2}$; this law
fits the data for $\C.C\ge 0.1$ and is not in contradiction with the 
(noisy) data even for $\C.C=0.01$.}
\label{pvsn}
\end{figure}

Since our arguments are not rigorous and the result quite 
surprising, we will examine the assumptions using an additional
consideration. From Eqs. (\ref{D2}) and (\ref{Dineq}) follows
that $D_2=1$, and from Eq. (\ref{layernumber}) it then
follows that $p$ scales like
\begin{equation}
p\sim S^{1/2} \ , \quad \C.C>\C.C_{\rm cr}  \ .
\end{equation}
This prediction is examined directly in Fig. \ref{pvsn}.
We see that it is obeyed extremely well for all the values
of $\C.C\ge 0.1$, and it is not in contradiction with the data
even for $\C.C=0.01$. We therefore cannot exclude the possibility
that $\C.C_{\rm cr} =0$. 
\begin{figure}
\centering
\includegraphics[width=.33\textwidth]{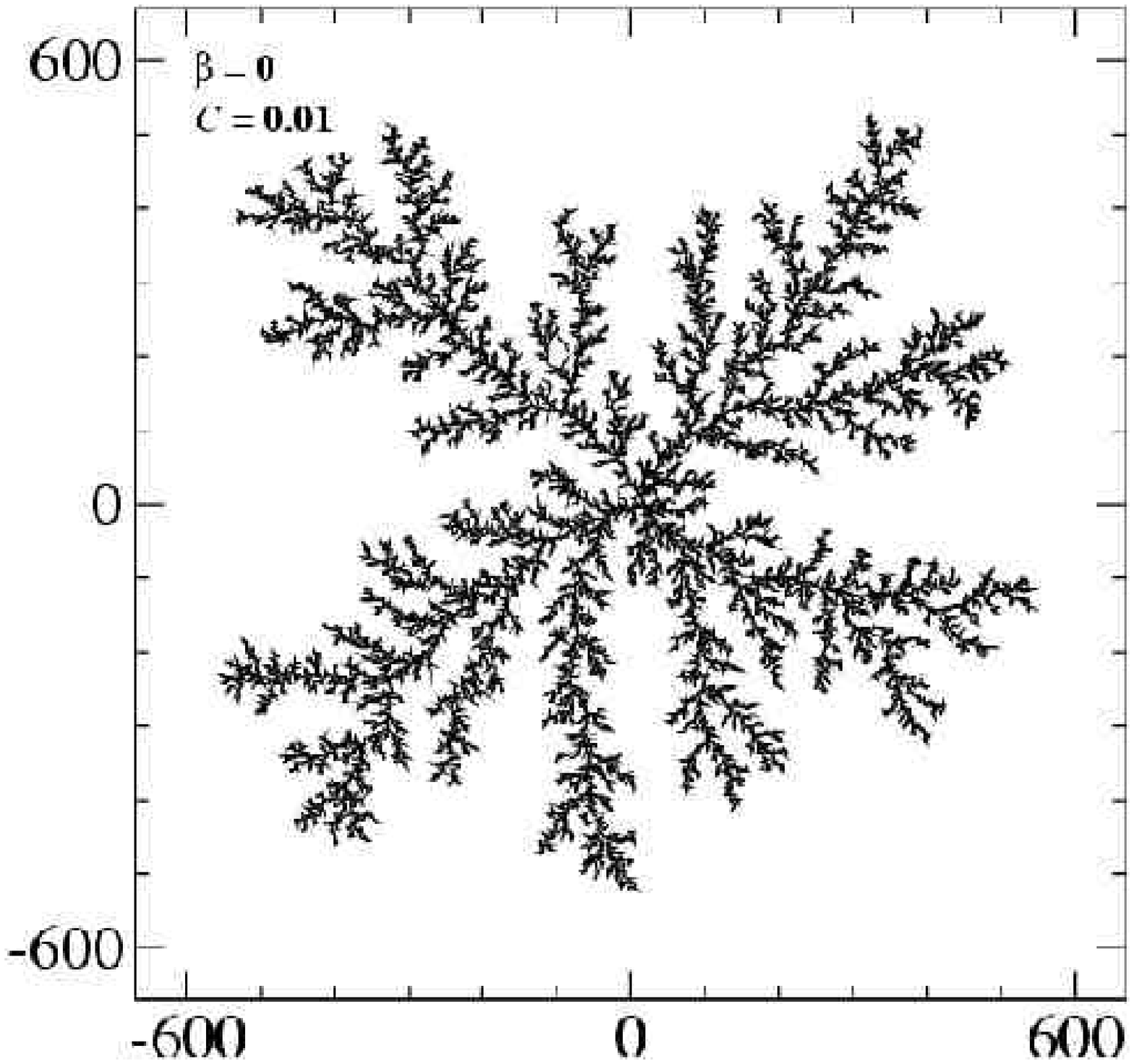}
\includegraphics[width=.33\textwidth]{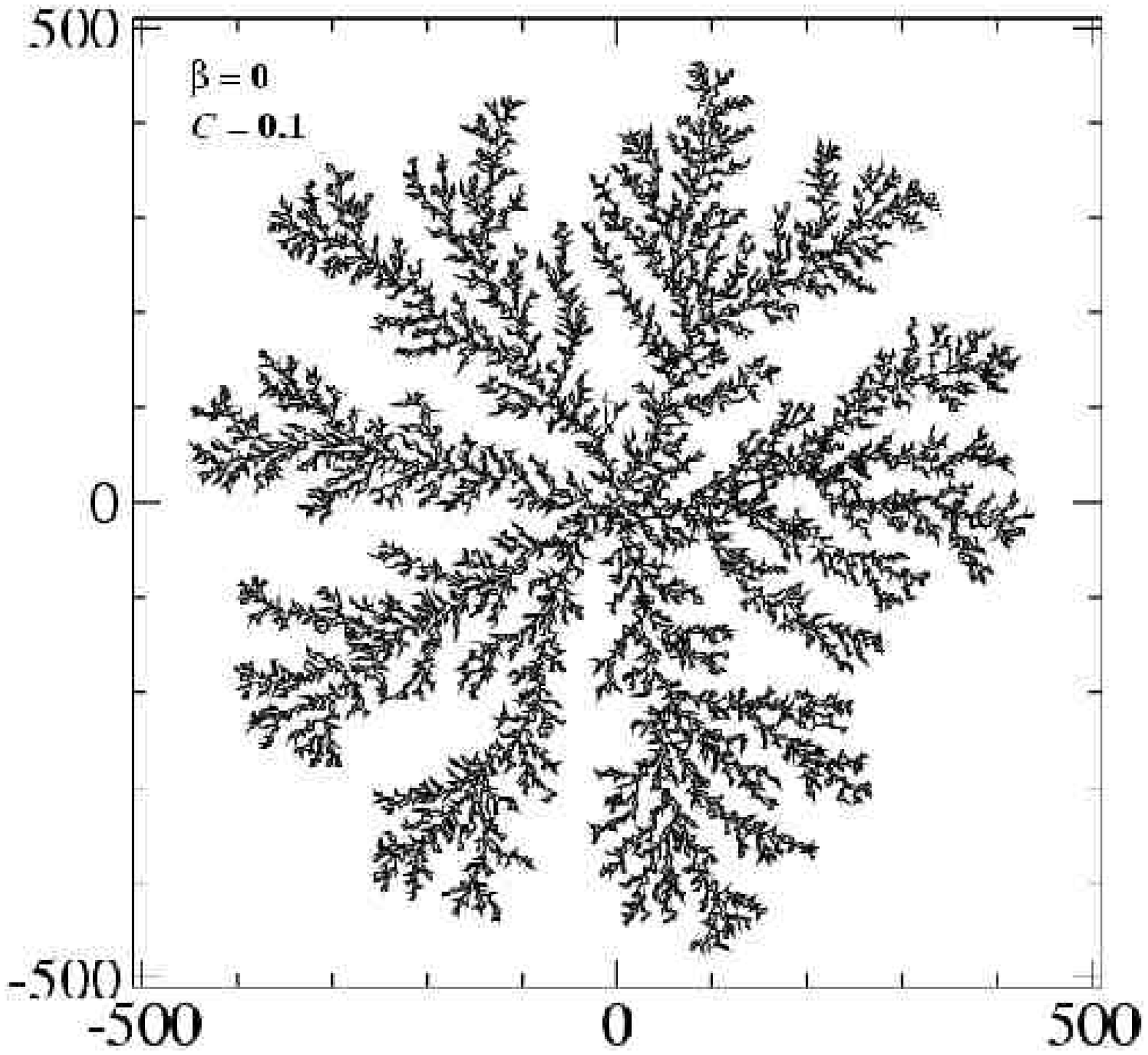}
\includegraphics[width=.33\textwidth]{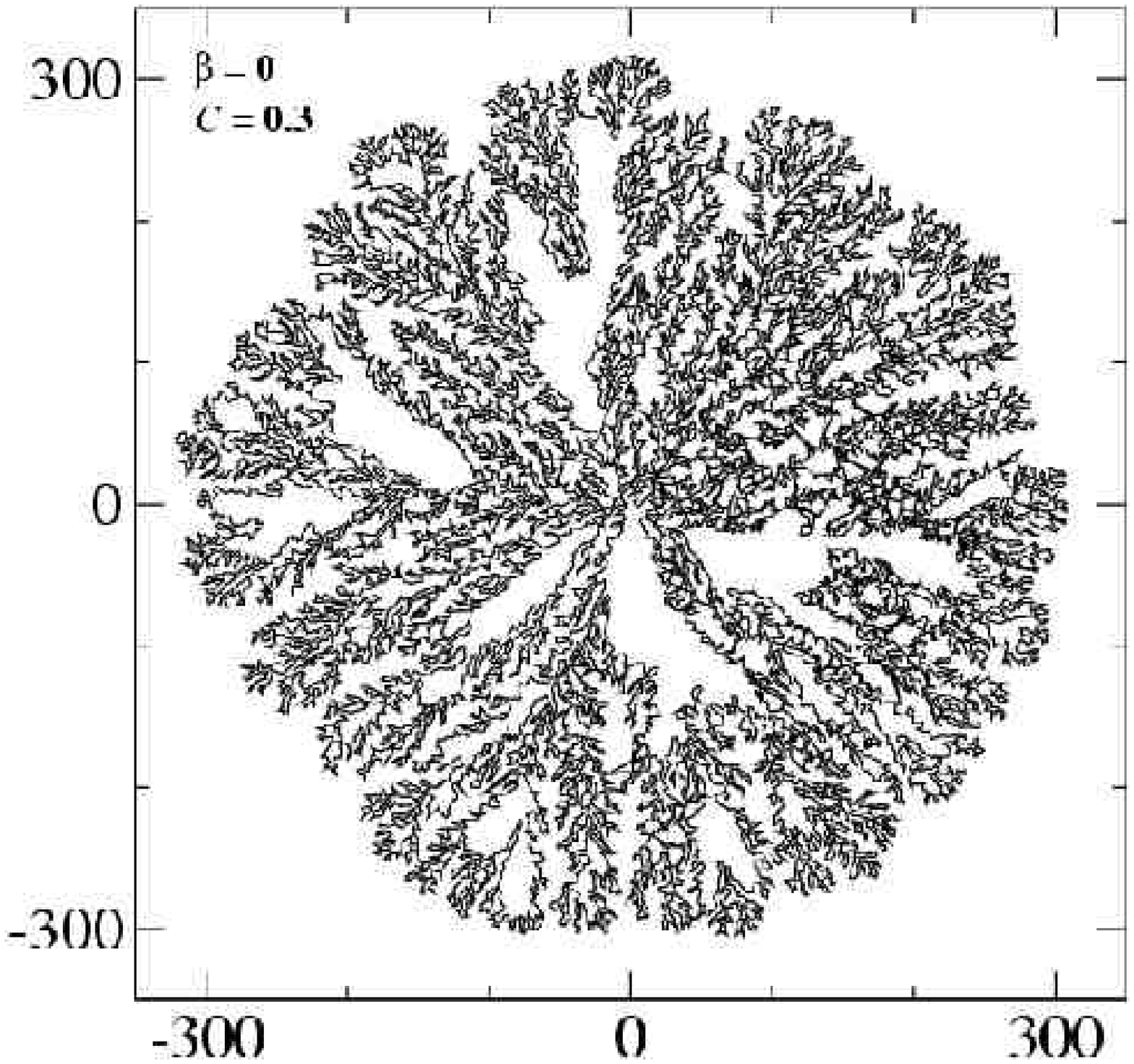}
\includegraphics[width=.33\textwidth]{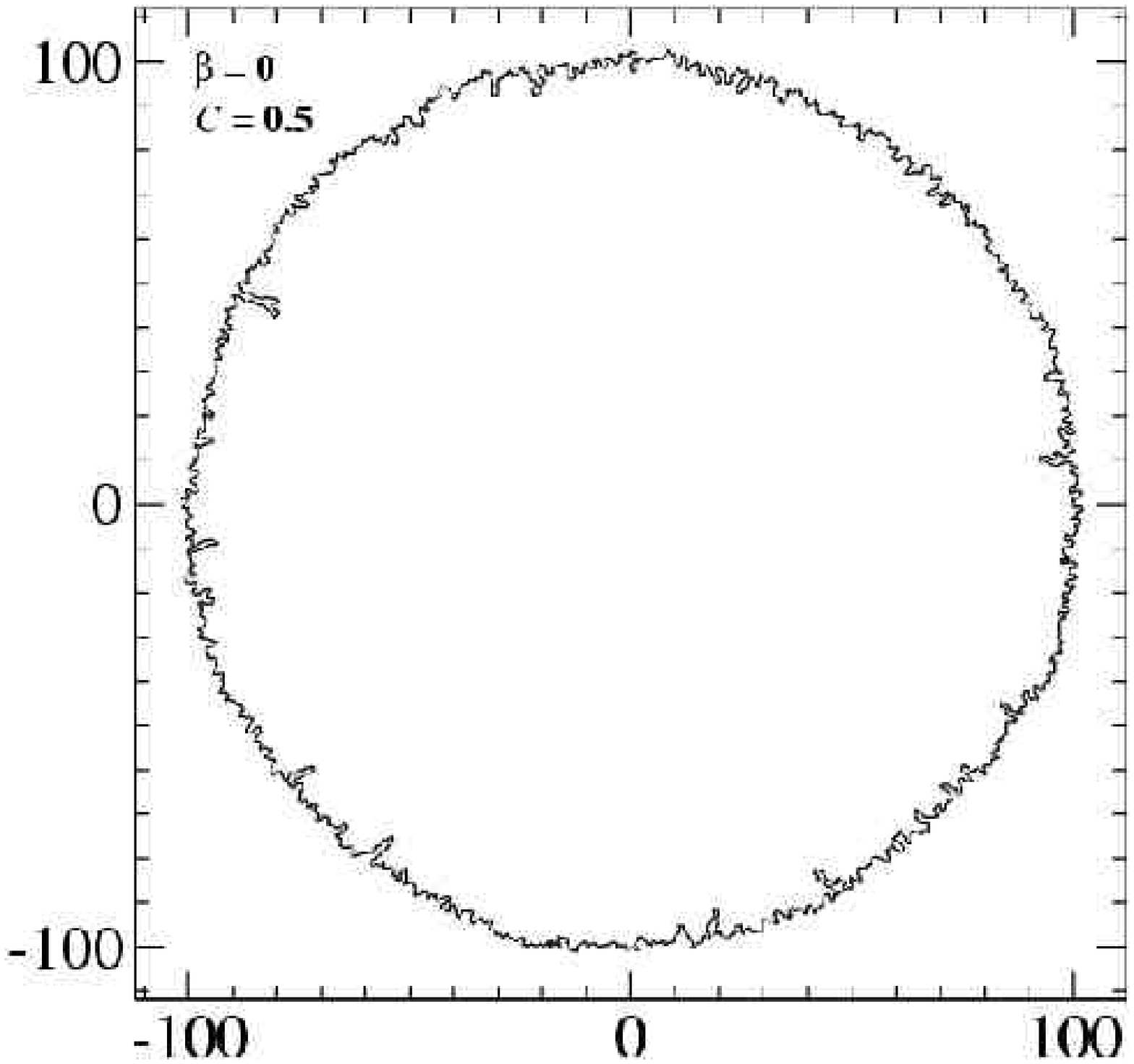}
\caption{Clusters for $\beta=0$. Panels a-d: ${\cal C}=0.01,
0.1,0.3$ and 0.5 respectively. Note the areas significantly larger than the UV-cutoff
$\lambda_0$ which appear already for ${\cal C}=0.01$.}
\label{clusters0}
\end{figure} 

To gain intuition to the meaning of this result
we show in Fig.
\ref{clusters0} the actual growth patterns for $\beta=0$ and ${\cal C}=0.01$, 0.1, 0.3
and 0.5 respectively. To plot these figures we find {\em all}
the exposed branch cuts on the unit circles which are associated with
the bumps added in the growth process (see \cite{01BDP} for details).
Then we plot the image of all these points under the conformal map and connect
them by lines. Thus we are guaranteed that what is plotted is the actual
contour of the growth pattern, of the image of the unit circle in the
mathematical domain, with all the fjords fully resolved. 
We see that even with the lowest value of
${\cal C}$ the branches appear to gain substance as they grow, having a width 
which is larger
than $\sqrt{\lambda_0}$ (the typical corrugation of the interface). 
Consequently it is not impossible that $D=2$ even for the lowest
values of $\C.C>0$. If this is so, it is NOT due to
the existence of an ultraviolet cutoff, but due to the finiteness of ${\cal C}$.
With $\C.C=0$ (the DLA limit) the serial algorithm favors strongly
truly fractal patterns. The parallel growth algorithm with finite $\C.C$
squeezes more substance into the fjords, reducing that tendency.
For higher values of ${\cal C}$ it becomes obvious that the 
growth patterns are 2-dimensional, and for ${\cal C}=0.5$
the pattern grows like a roughened disk. The main conclusion of this analysis
is that we certainly cross somewhere along the line
$\beta=0$ into growth patterns that are 2-dimensional. Whether or not
the critical value of $\C.C$ is finite or zero cannot be determined
by numerics alone.
\begin{figure}
\centering
\includegraphics[width=.33\textwidth]{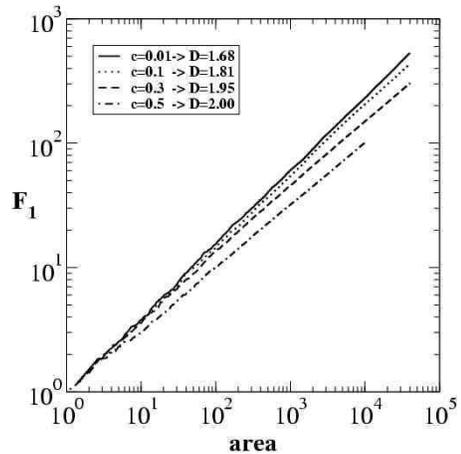}
\caption{The first Laurent coefficient $F_1^{(n)}$ 
as a function of the area for $\beta=0$
and ${\cal C}=0.01, 0.1,0.3$ and 0.5. The fractal 
dimension $D$ is obtained for the slope
via $F_1^{(n)} \sim \sqrt{\lambda_0} (S/\lambda_0)^{1/D}$ }
\label{firstlaurent0}
\end{figure} 

If we accept the possibility that even the lowest values of $\C.C$
are associated with growth patterns that are 2-dimensional, then
we should stress that standard ways of estimating the
dimension of these clusters, especially for the lowest value of $\C.C$,
may fail to discover this fact. For example,
we can compute $F_1^{(n)}$ and then, using Eq.(\ref{scaleF1}), attempt
to extract the dimension from log-log plots or $F_1^{(n)}$ against
$S$. This method works very well for the fractal case, but it 
does not appear to do so well for the cases at hand. In Fig. \ref{firstlaurent0}
we show such log-log plots for all the clusters of Fig. \ref{clusters0}.
We see that even with 100 000 particles the dimension estimate is
way below the suspected $D=2$, except for $\C.C=0.5$. In fact, any practitioner in the fractal
field would be happy to interpret the scaling obtained for 
$\C.C=0.01$ as an indication that it is in the same universality
class of DLA with dimension very close to $D=1.71$.  While
we cannot state confidently that for $\C.C=0.01$ the growth
pattern is 2-dimensional, we stress that the dimension estimates obtained from
log-log plots can be only taken as lower bounds on the true dimension, and these may
not be very sharp.

\begin{figure}
\centering
\includegraphics[width=.33\textwidth]{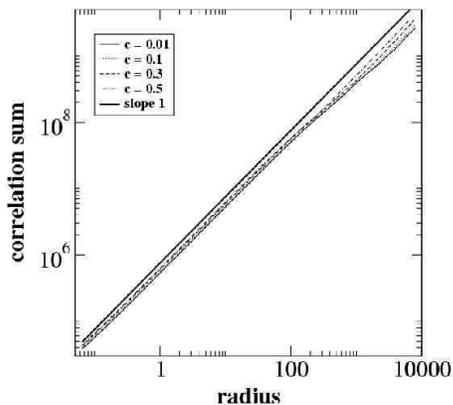}
\caption{The correlation dimension $D_2$ for 
$\beta=0$ and ${\cal C}=0.01,
0.1,0.3$ and 0.5. The thick line has slope 1, 
indicating that $D_2=1$ and therefore $D=2$
for all shown ${\cal C}$.}
\label{corr_dim}
\end{figure}

A possibly better way to measure the dimension would be through the result
(\ref{D2}) when it holds. We have very good methods to determine the correlation
dimension $D_2$, going back to the Grassberger-Procaccia
algorithm \cite{83GP}. To this aim we choose randomly $m=100 000$ points
$\{\theta_i\}_{i=1}^m$, and compute their positions on the
interface of the cluster $z_i=\Phi^{(n)}(e^{i\theta_i})$. Next 
we compute the correlation integral
\begin{equation}
C^{(2)}(r) =\sum_{i\ne j} \Theta(|z_i-z_j|-r) \ , \label{corint}
\end{equation}
where $\Theta(x)$ is the step function, being 1 for $x\le 0$
and 0 for $x>0$. The correlation integral is known to scale according
to 
\begin{equation}
C^{(2)}(r)\sim r^{D_2} \ . \label{Cscale}
\end{equation}
In Fig. \ref{corr_dim} we display this object in a log-log plot
as a function of $r$. All the values of $\C.C$ agree with a
correlation dimension of $D_2=1$, as can be seen from the
plots at small scales. For those values of $\C.C$ for which 
Eq. (\ref{D2}) is correct this leads to the aforementioned result $D=2$. 

\subsection{The case $\beta=2$ and ${\cal C}>0$}
\begin{figure}
\centering
\includegraphics[width=.33\textwidth]{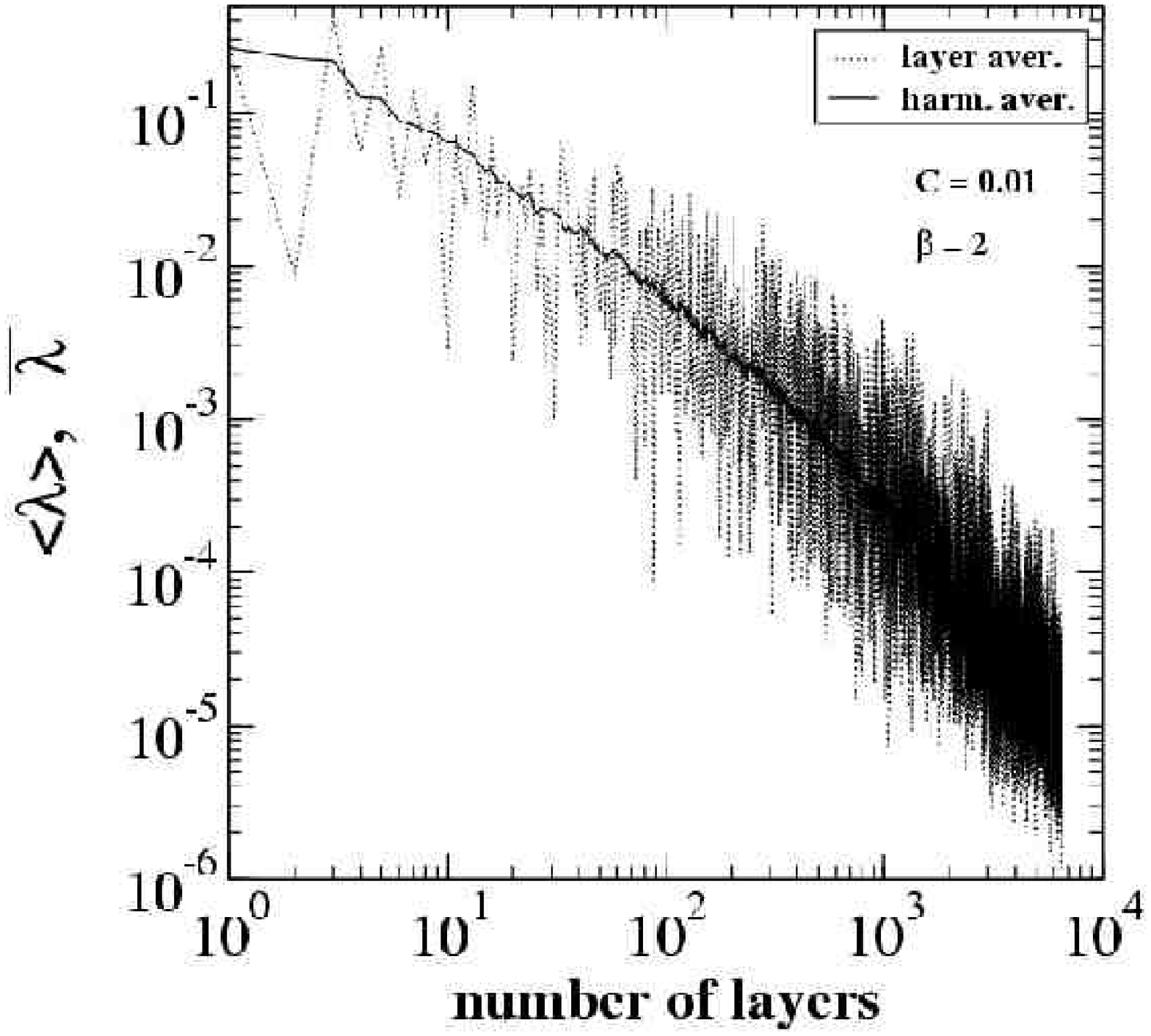}
\includegraphics[width=.33\textwidth]{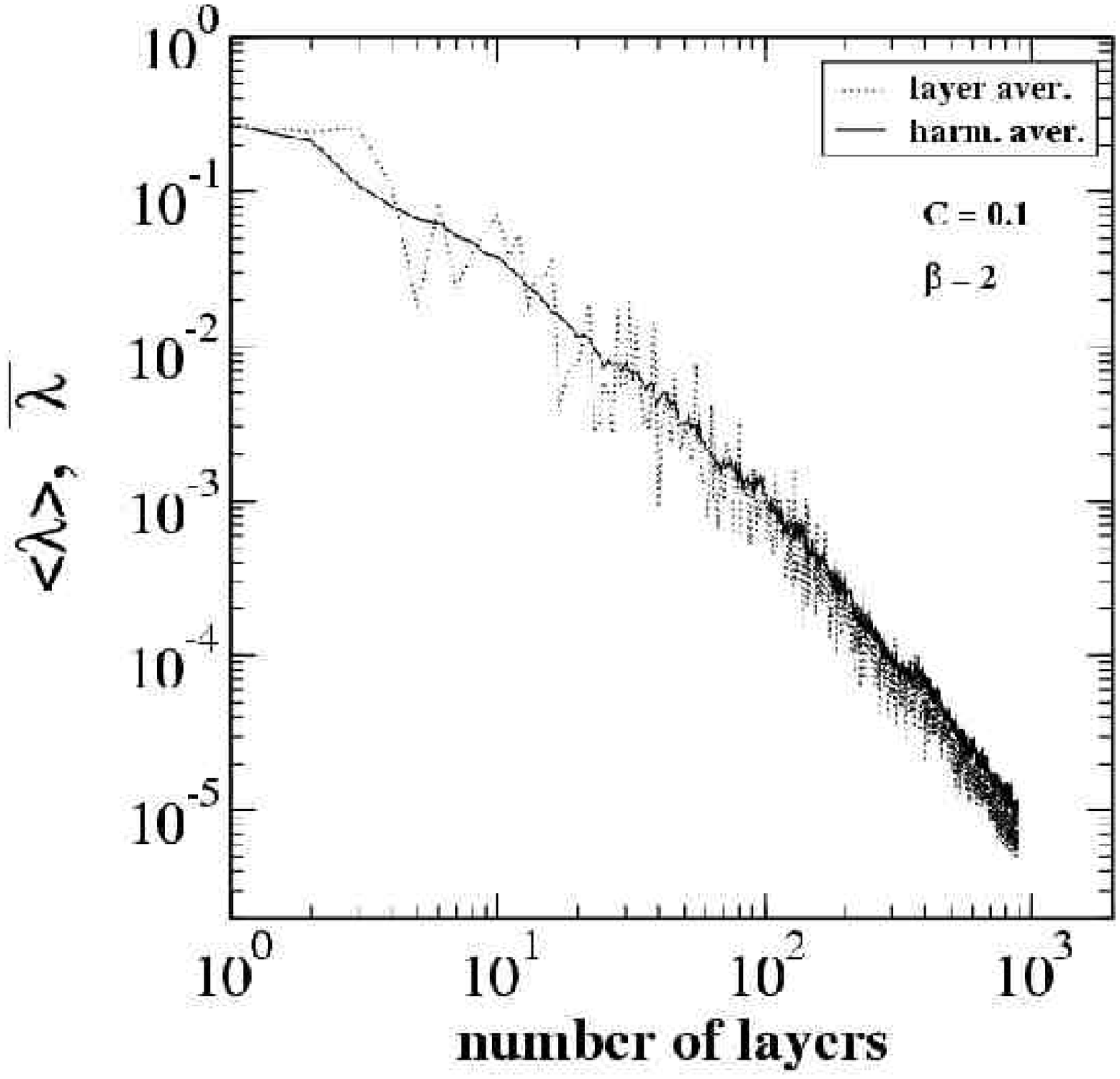}
\includegraphics[width=.33\textwidth]{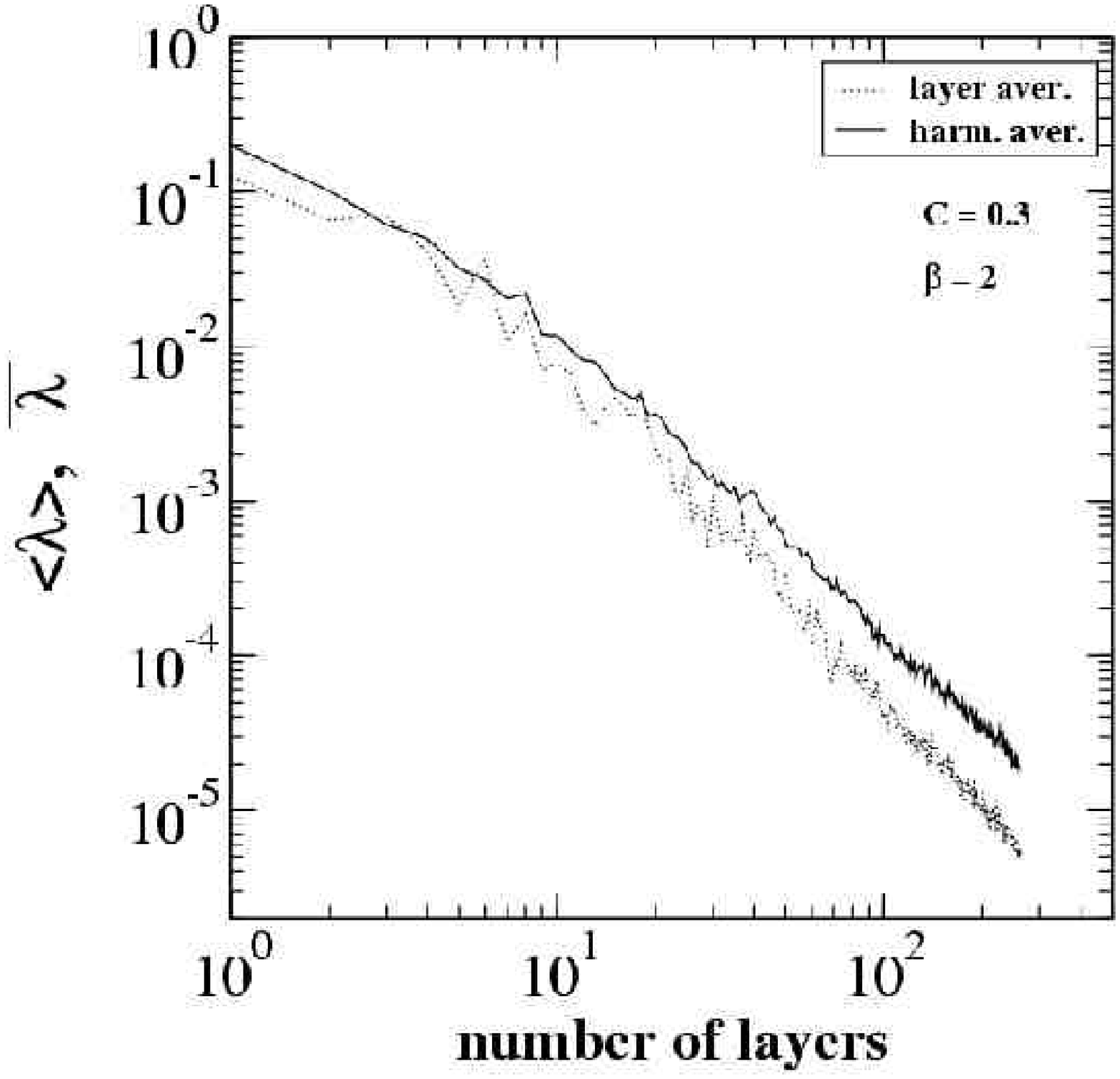}
\includegraphics[width=.33\textwidth]{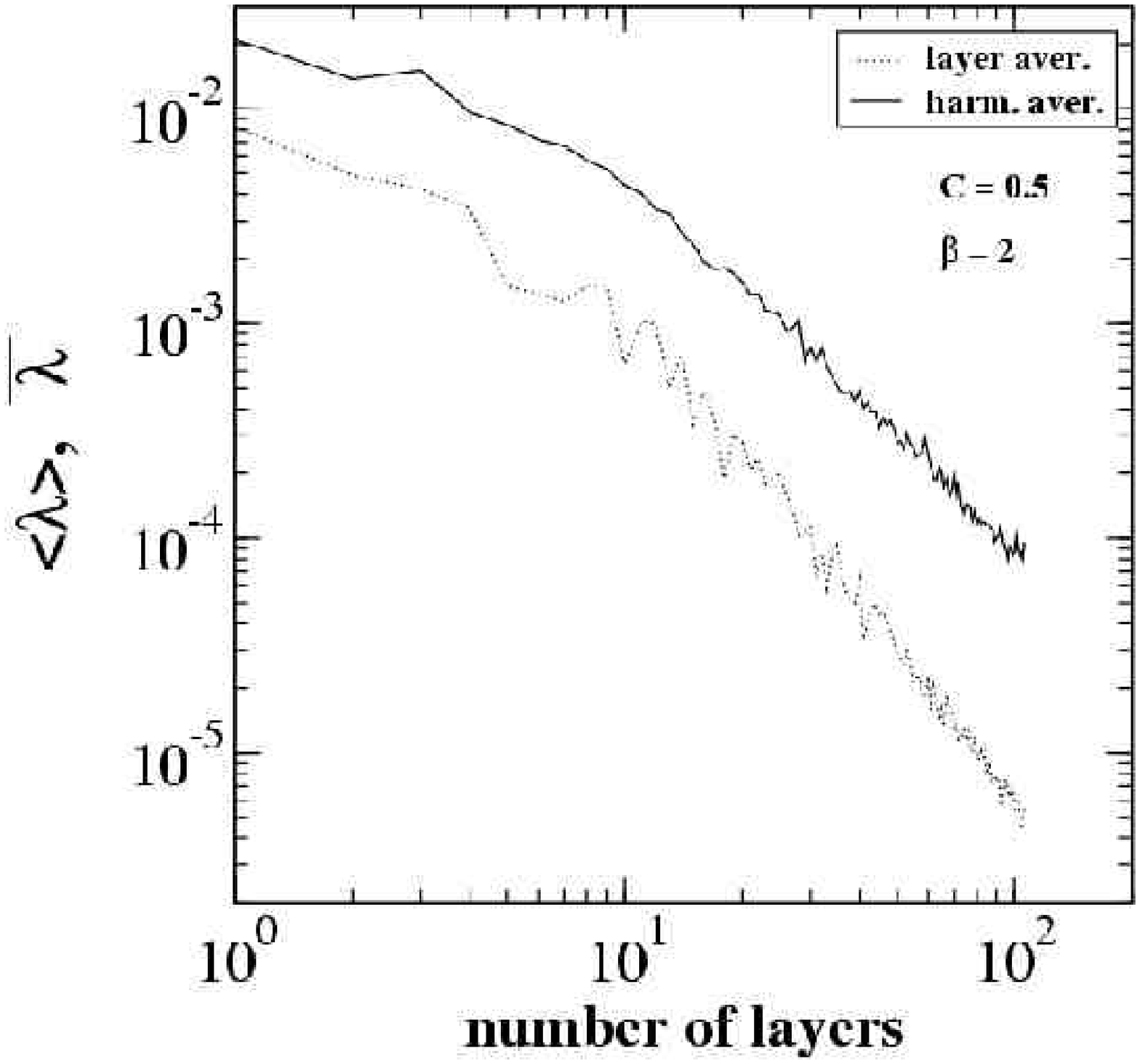}
\caption{Layer and harmonic averages of $\lambda_n$ as a function of the
number of layers, for $\beta=2$. Panels a-d: ${\cal C}=0.01,
0.1,0.3$ and 0.5 respectively.}
\label{averages21}
\end{figure}

\begin{figure}
\centering
\includegraphics[width=.33\textwidth]{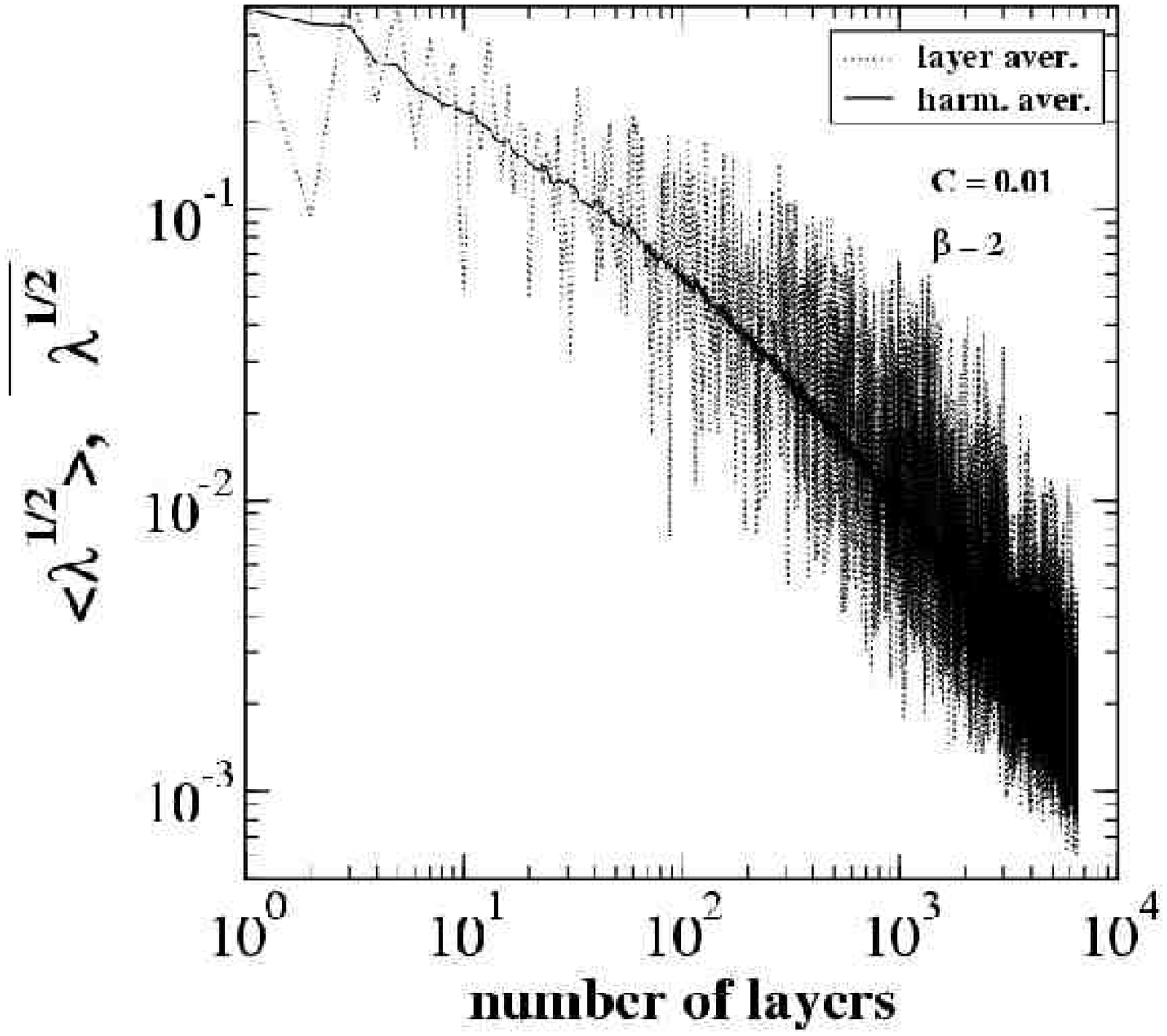}
\includegraphics[width=.33\textwidth]{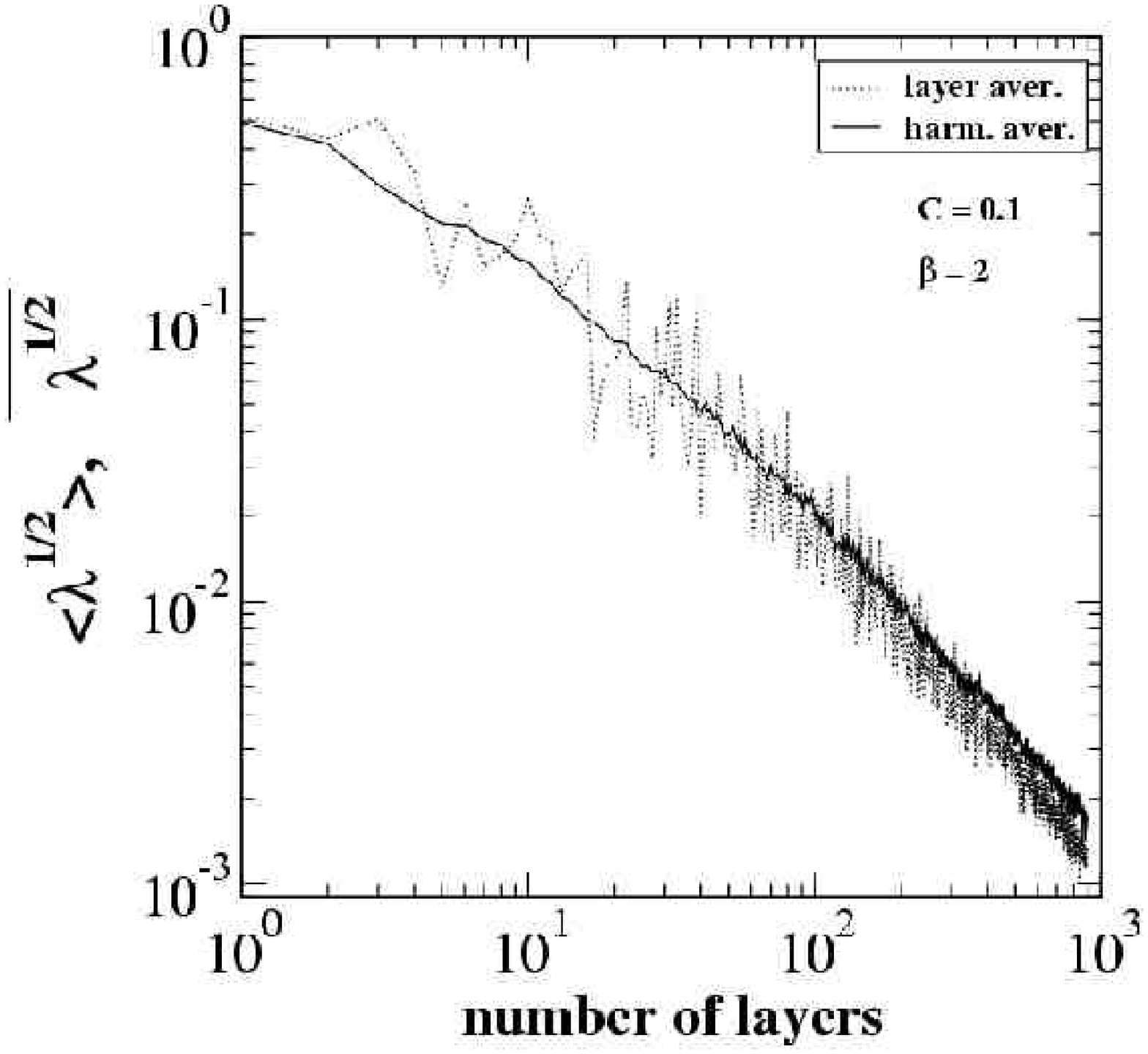}
\includegraphics[width=.33\textwidth]{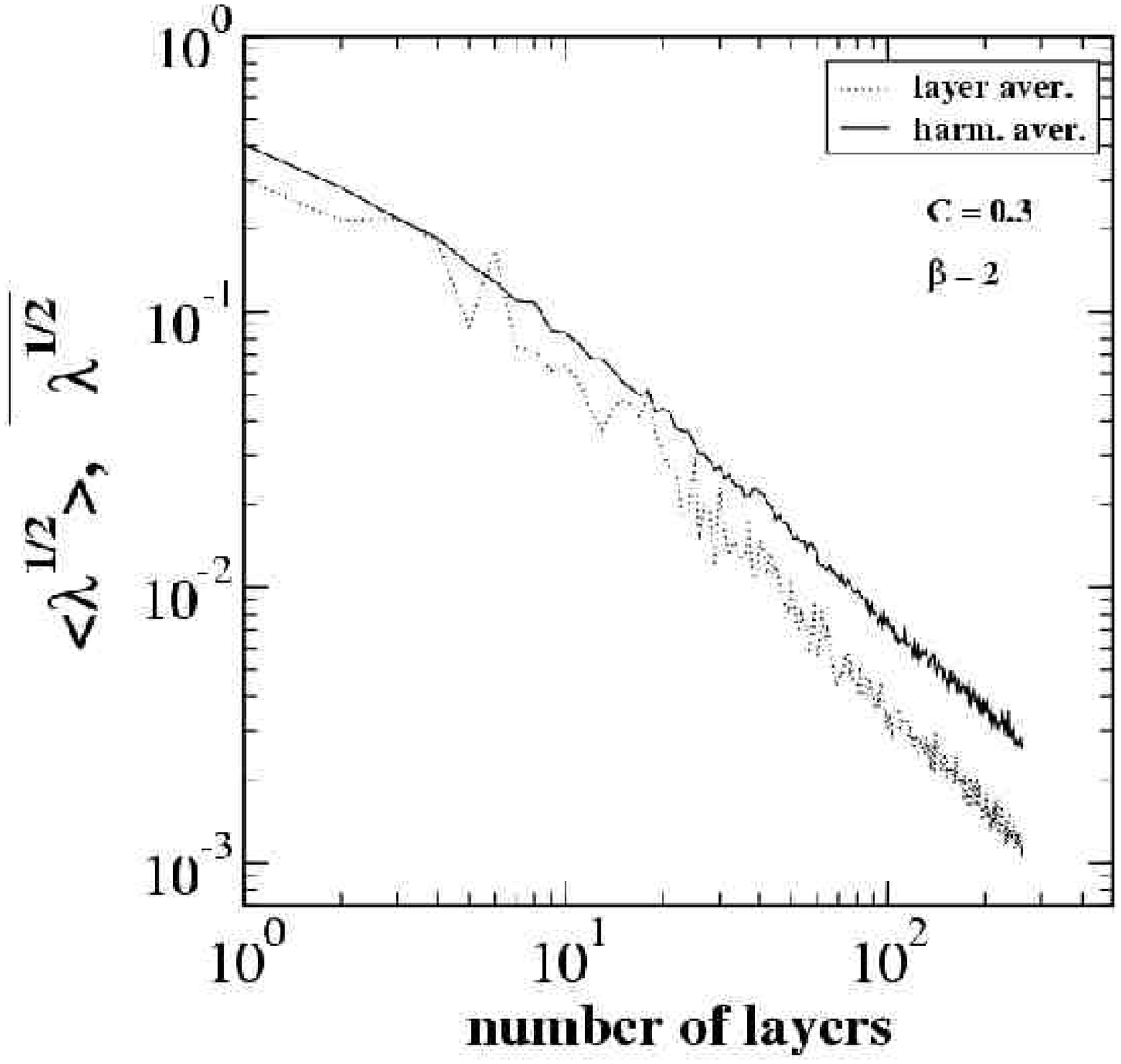}
\includegraphics[width=.33\textwidth]{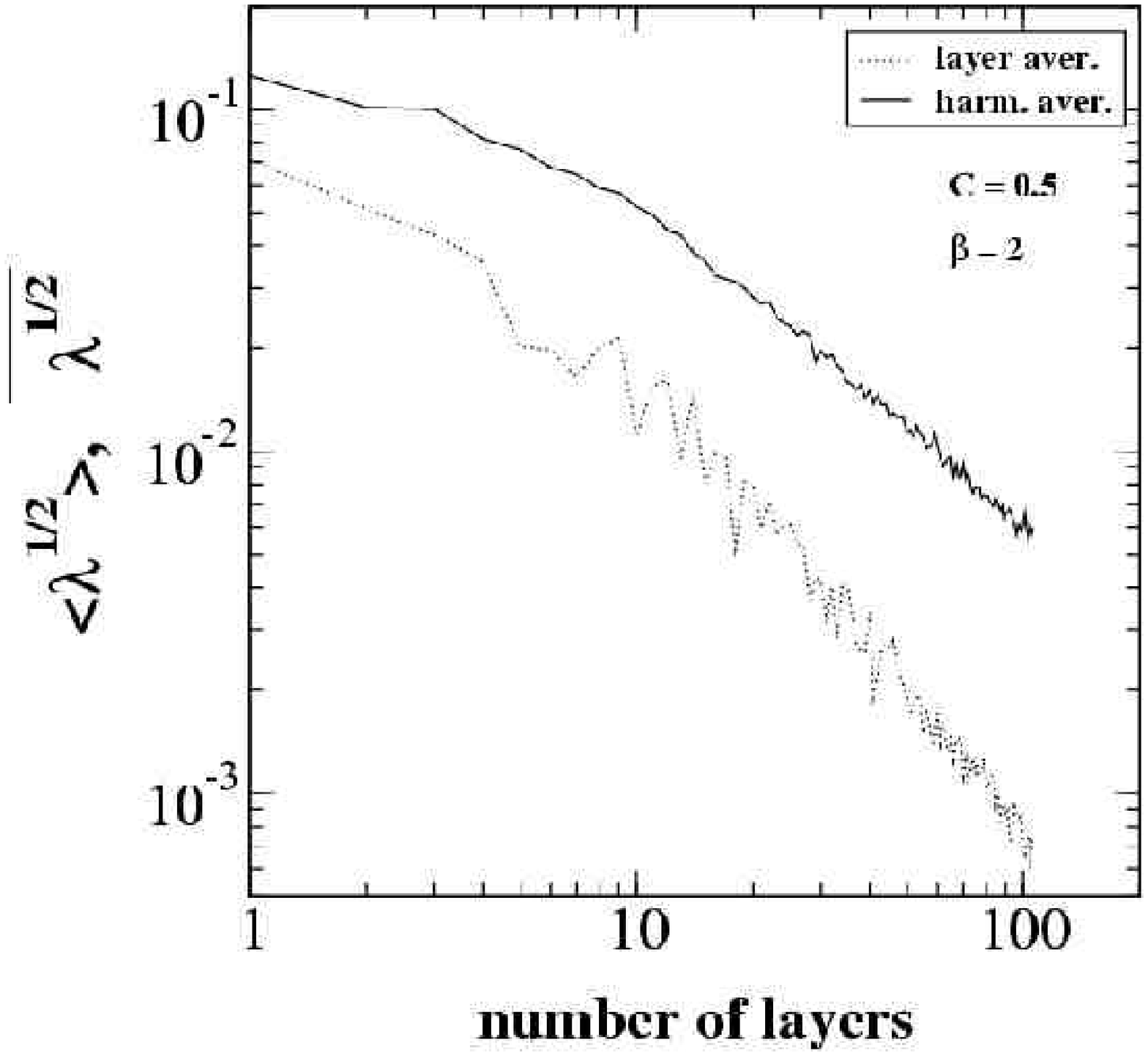}
\caption{Layer and harmonic averages of $\lambda^{0.5}_n$ as a function of the
number of layers, for $\beta=2$. Panels a-d: ${\cal C}=0.01,
0.1,0.3$ and 0.5 respectively.}
\label{averages205}
\end{figure}

The next interesting family of growth patterns that we
focus on is obtained for $\beta=2$ and $\C.C>0$, with
Laplacian Growth expected to be realized for $\C.C=1$.
We find that
for $\beta >0$ the numerics does not support the scaling relation
(\ref{averages}). In Figs. (\ref{averages21}) and (\ref{averages205})
we show the layer and harmonic averages for $\beta=2$, and it is obvious that in
this case 
 \begin{equation}
 \label{averages2}
 \overline{\lambda_n^q} \le  \langle \lambda_n^q \rangle
 \end{equation}
in the scaling sense.  

\begin{figure}
\centering
\includegraphics[width=.33\textwidth]{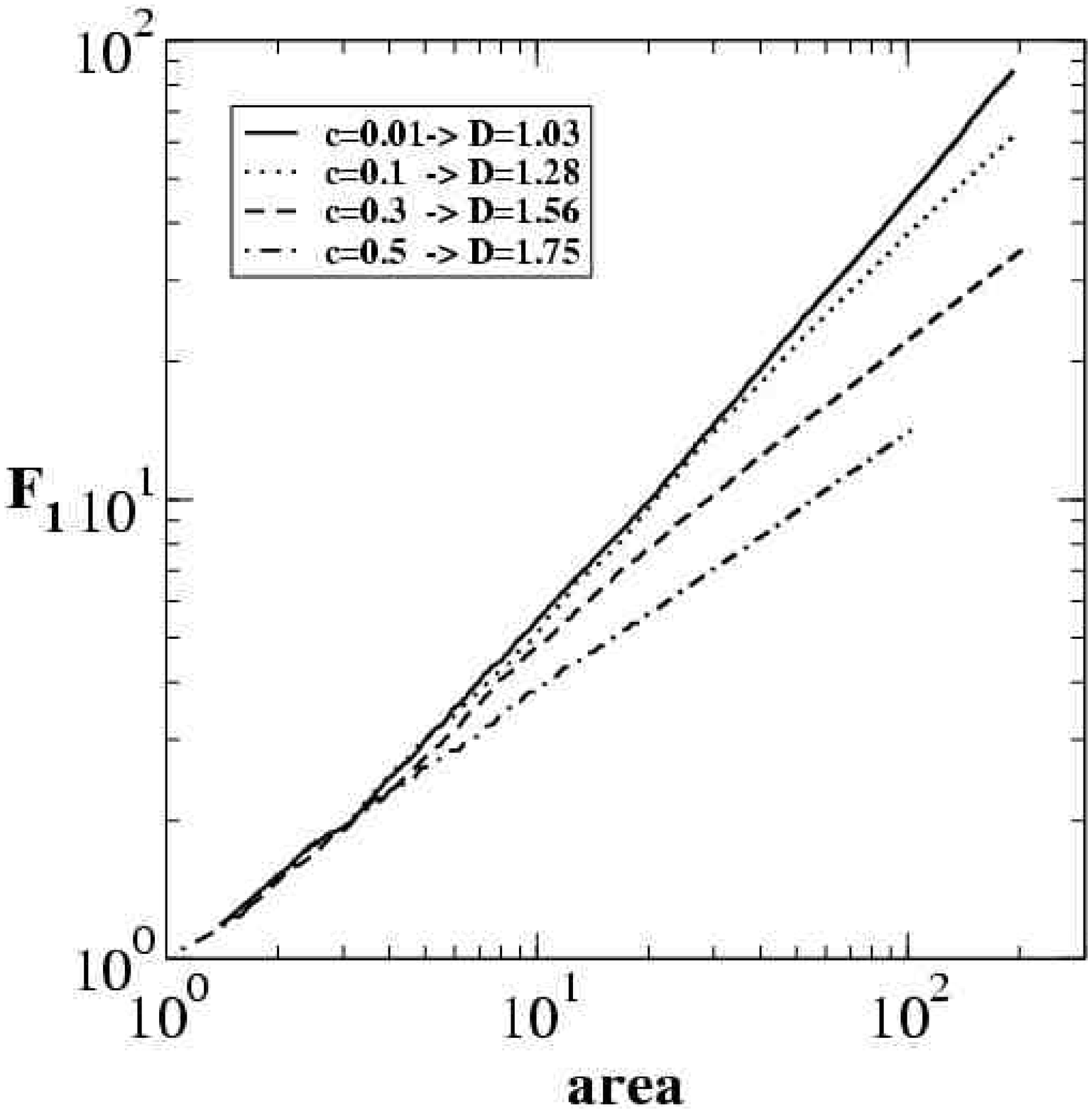}
\caption{The first Laurent coefficient $F_1^{(n)}$ as a function of the area for $\beta=2$
and ${\cal C}=0.01, 0.1,0.3$ and 0.5. The fractal dimension $D$ is obtained for the slope
via $F_1^{(n)} \sim \sqrt{\tilde\lambda_0} (S/\tilde\lambda_0)^{1/D}$ }
\label{firstlaurent2}
\end{figure} 

Once we have lost the scaling relation (\ref{averages}) we cannot argue that $D=2$ for 
any value of $\C.C>0$. We will find numerically that 
along the line $\beta=2$ we indeed find fractal patterns,
(and cf. the next section); nevertheless even along
this line there exists a transition to 2-dimensional patterns, 
albeit at a finite
and rather high value of $\C.C$. Next we want to estimate this value.
\section{Conjecture: Laplacian Growth is 2-Dimensional}
\begin{figure}
\centering
\includegraphics[width=.33\textwidth]{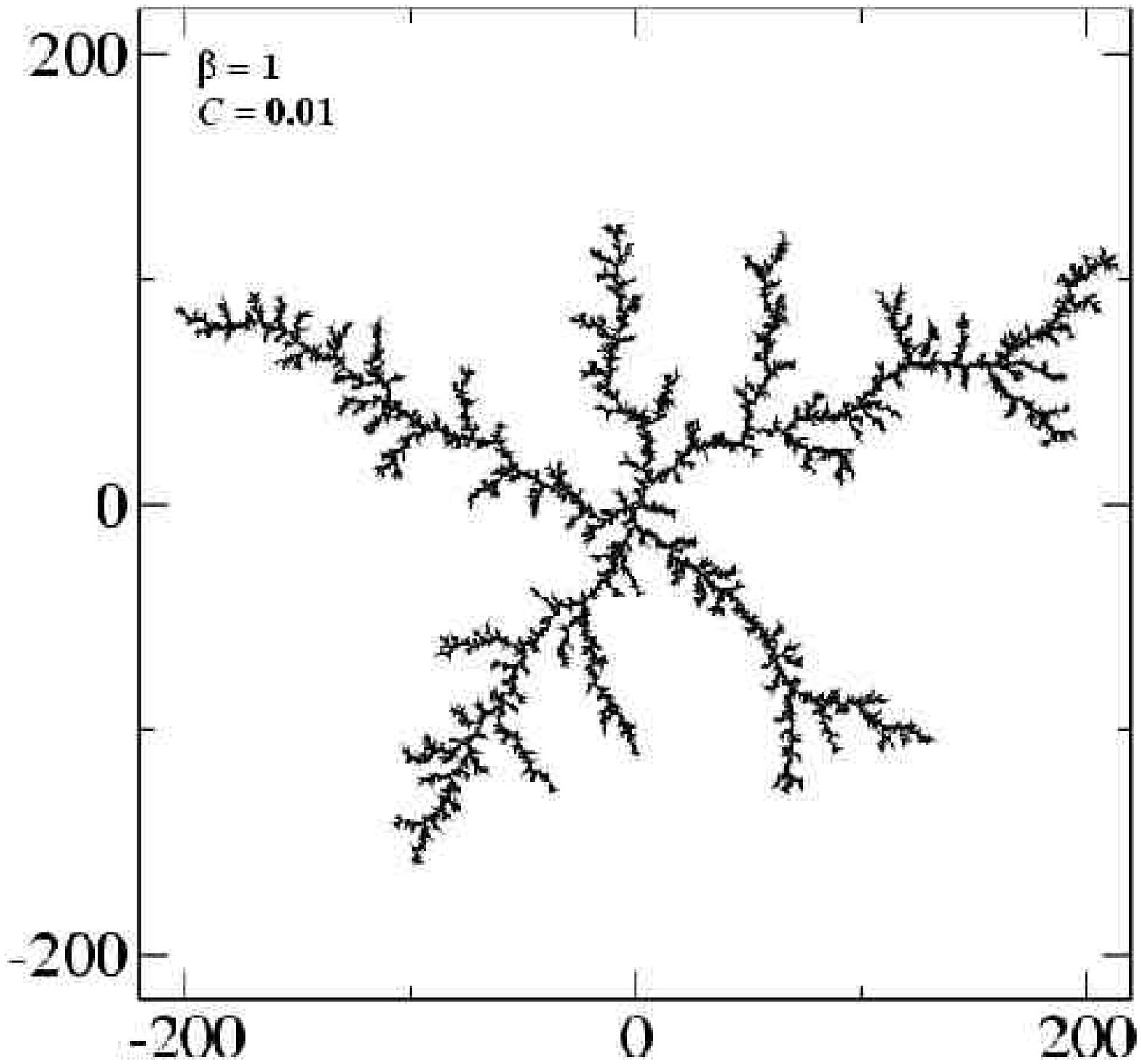}
\includegraphics[width=.33\textwidth]{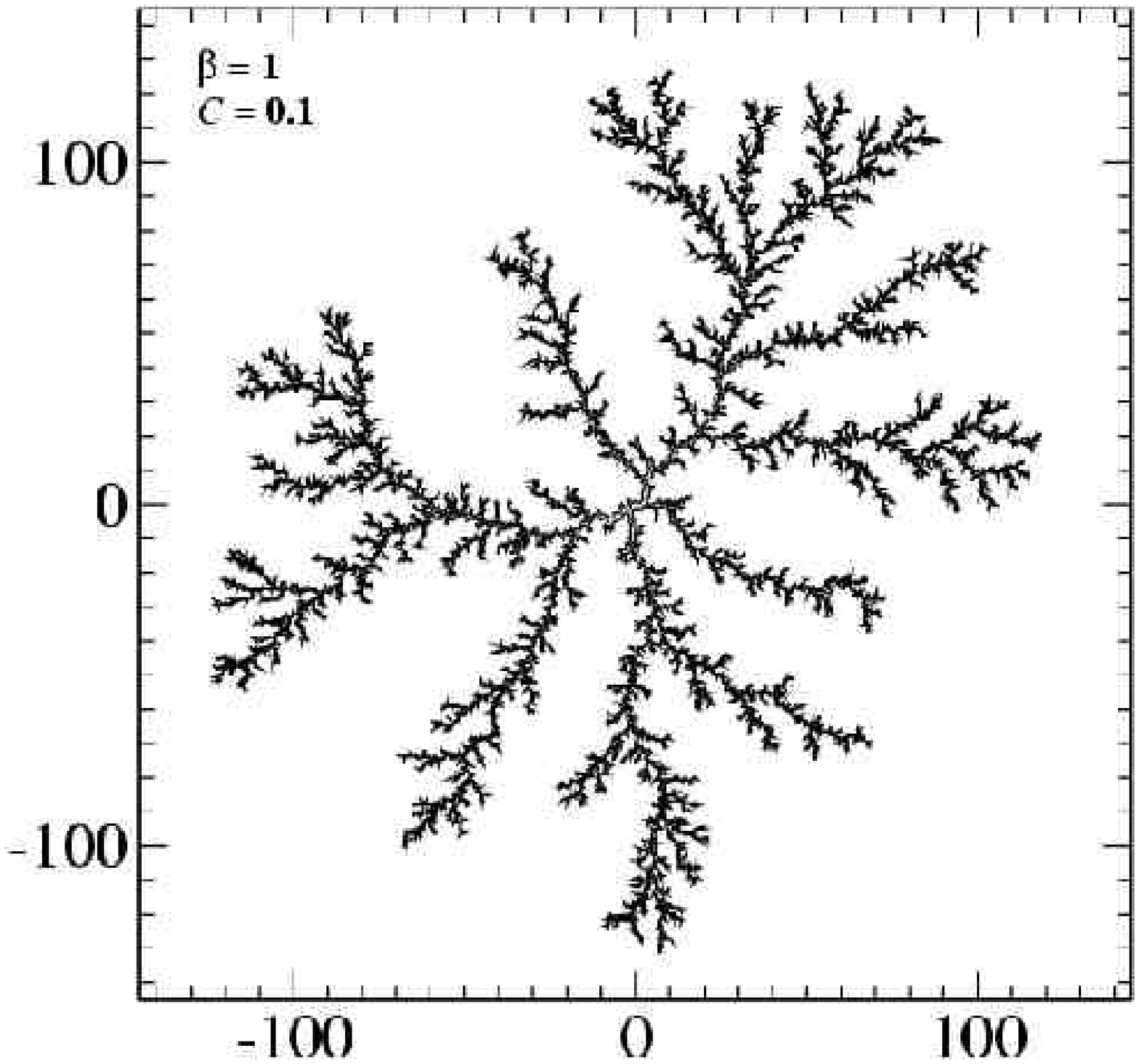}
\includegraphics[width=.33\textwidth]{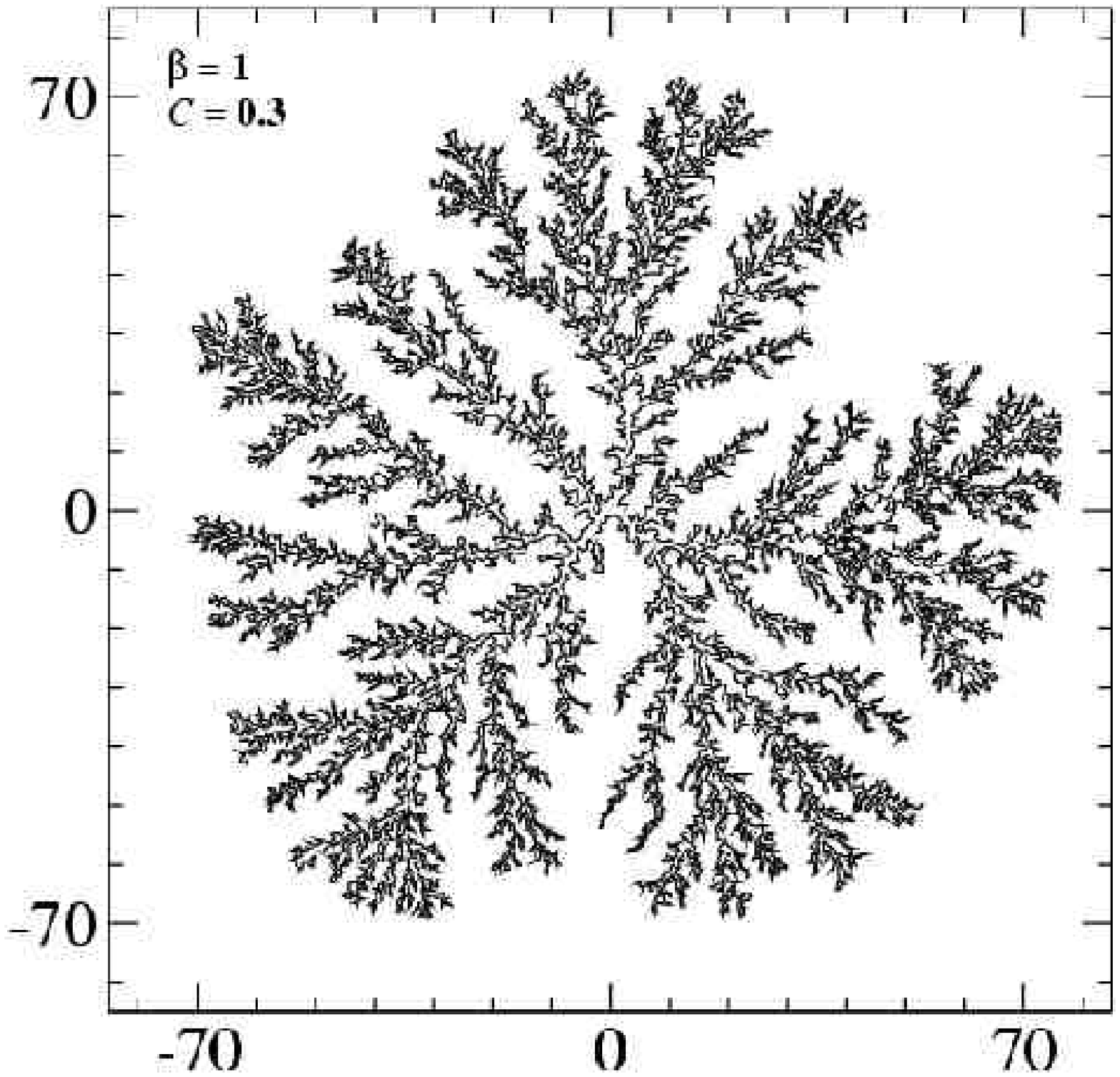}
\includegraphics[width=.33\textwidth]{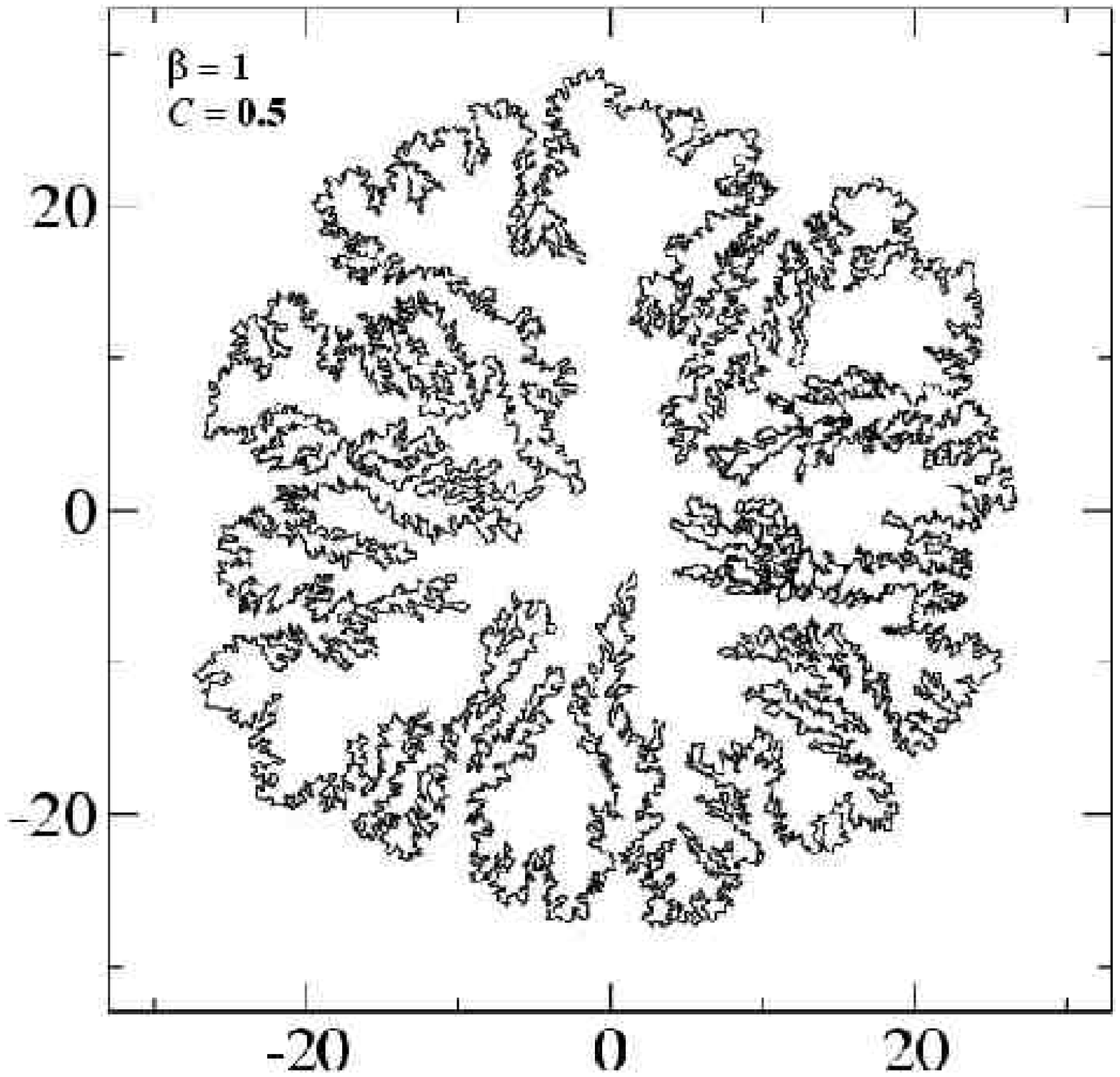}
\caption{Growth patterns for $\beta=1$. Panels a-d: ${\cal C}=0.01,
0.1,0.3$ and 0.5 respectively.}
\label{clustersbeta1}
\end{figure}

\begin{figure}
\centering
\includegraphics[width=.33\textwidth]{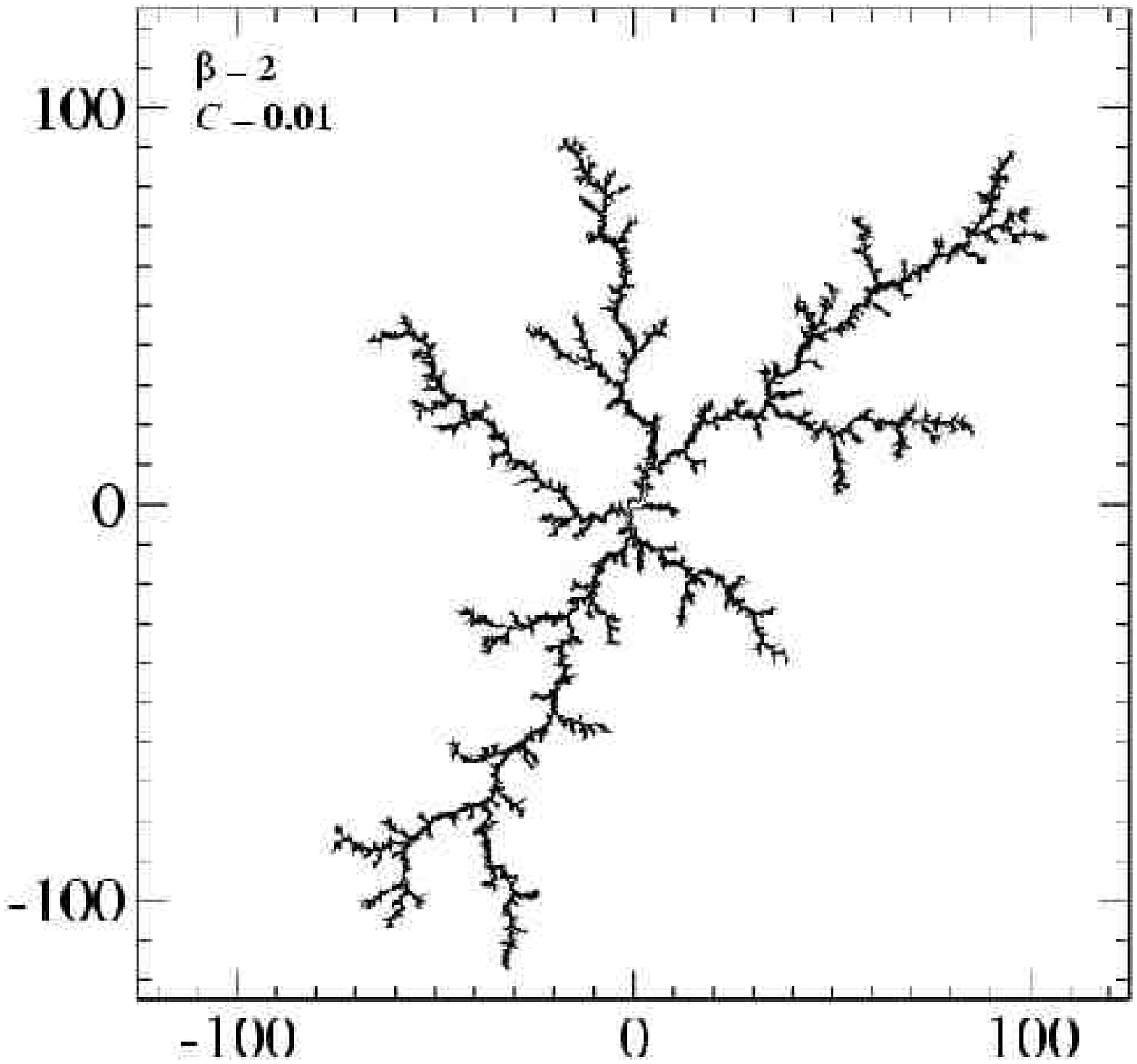}
\includegraphics[width=.33\textwidth]{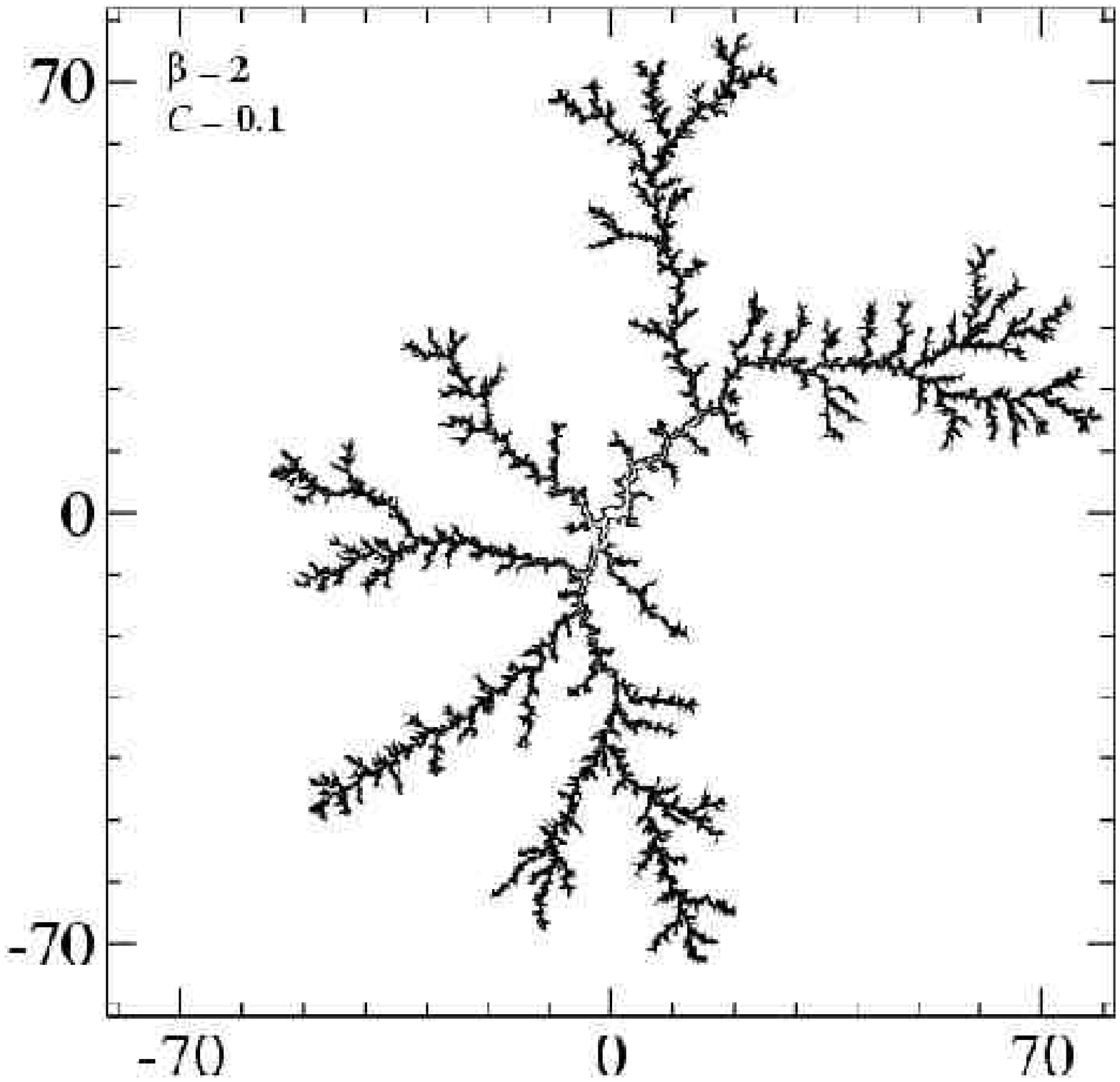}
\includegraphics[width=.33\textwidth]{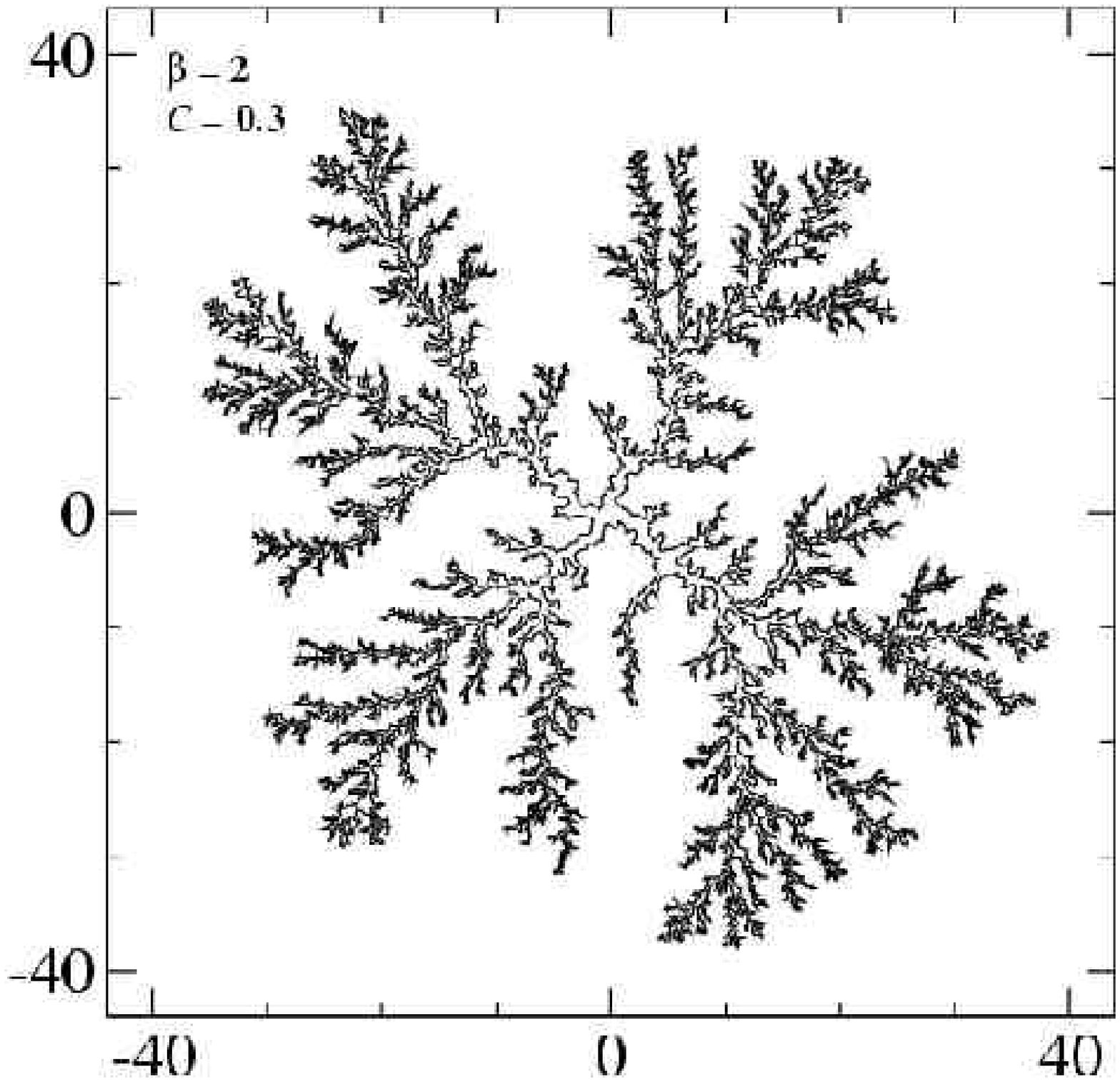}
\includegraphics[width=.33\textwidth]{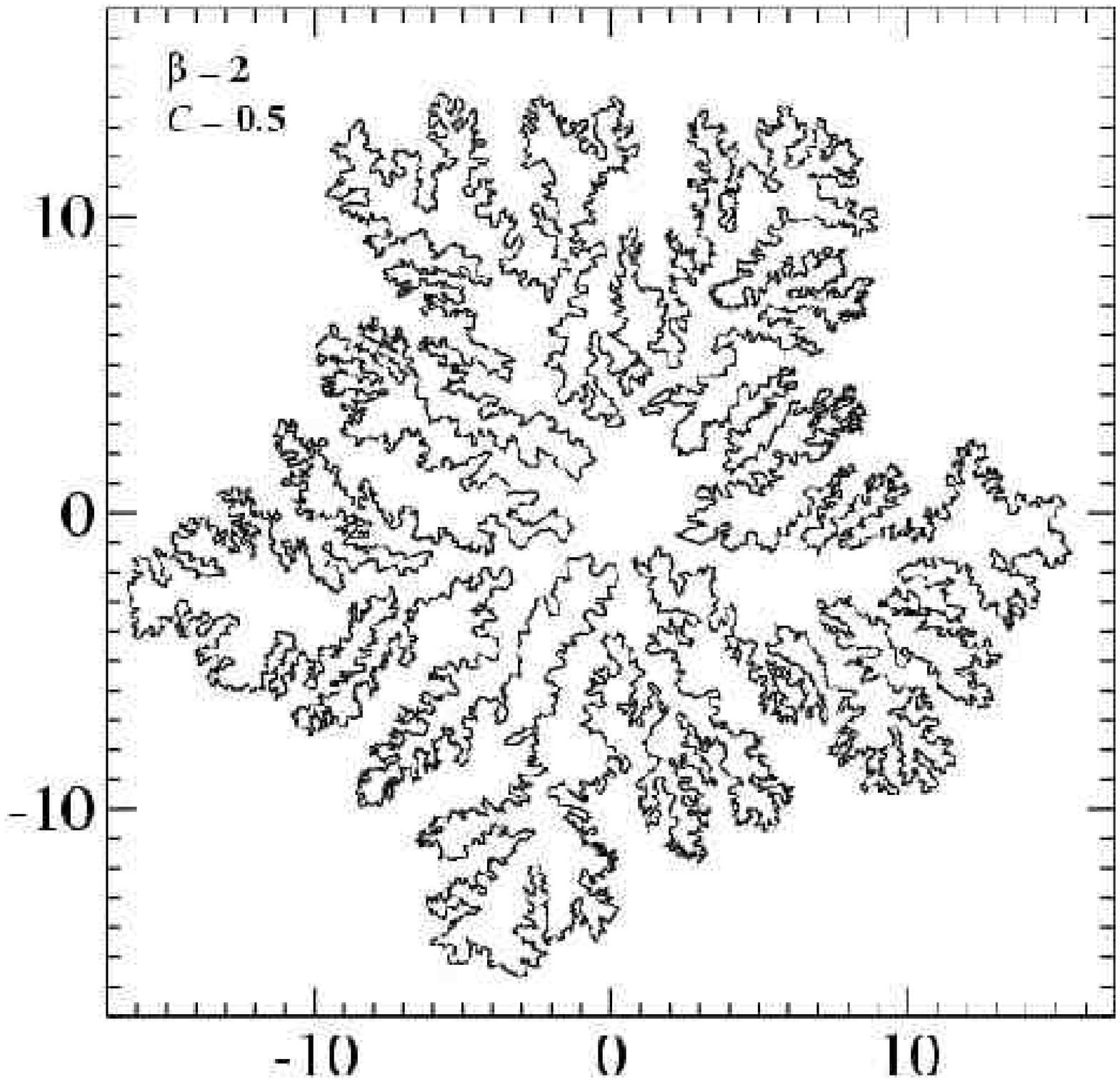}
\caption{Growth patterns for $\beta=2$. Panels a-d: ${\cal C}=0.01,
0.1,0.3$ and 0.5 respectively.}
\label{clusters2}
\end{figure}

In this section we motivate our conjecture that Laplacian
Growth patterns are not fractal patterns at all, but rather
patterns of dimension 2. We have to be a bit circumvent,
since as explained in \cite{01BDP}, we cannot directly run our
algorithm for the $\beta-\C.C$ model for values of $\C.C$
higher than about 0.65. The reason is that it becomes impossible
to fill up, by random selection of points on the
unit circle, a full layer of bumps on the physical interface.
\begin{figure}
\centering
\includegraphics[width=.33\textwidth]{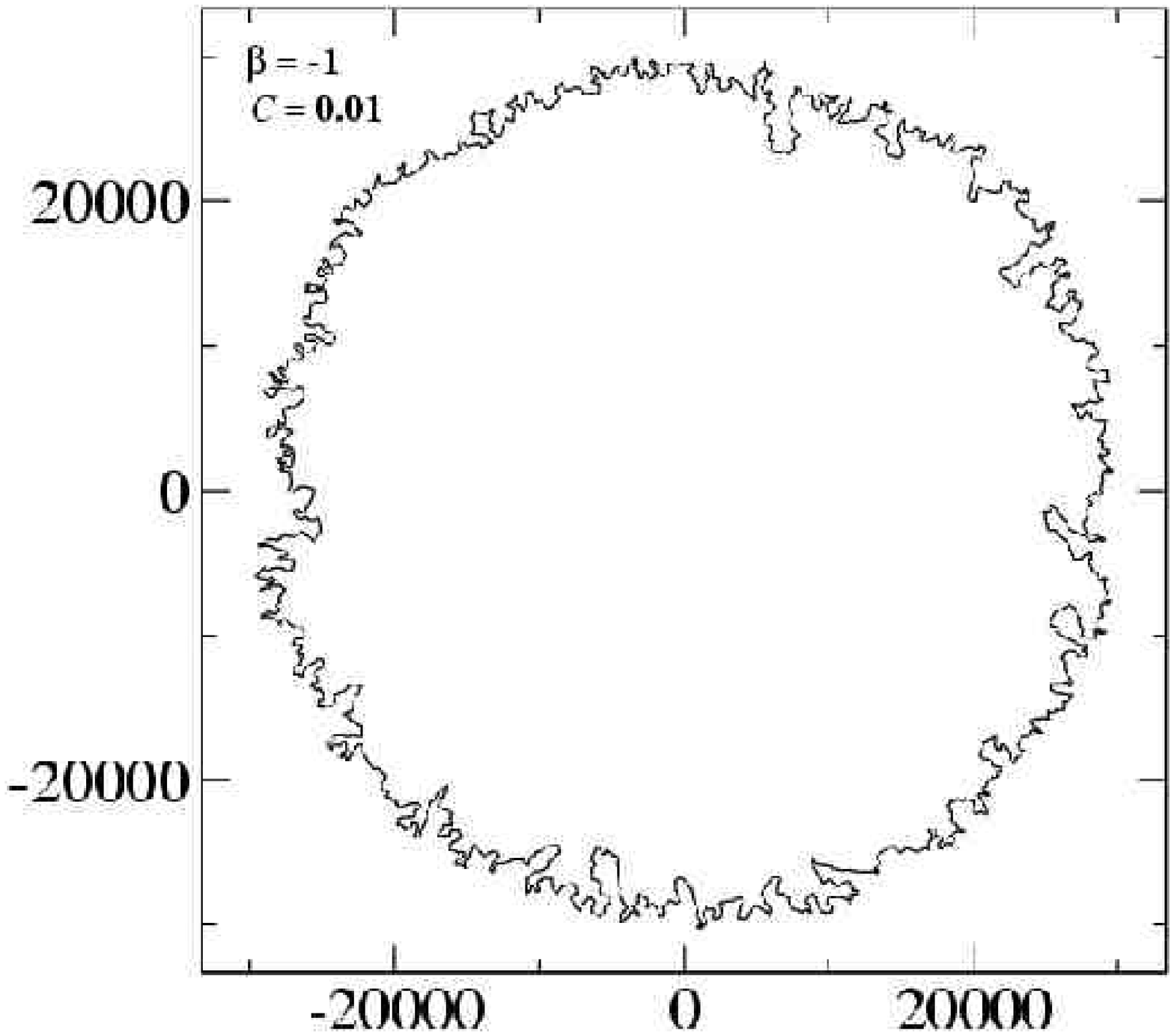}
\includegraphics[width=.33\textwidth]{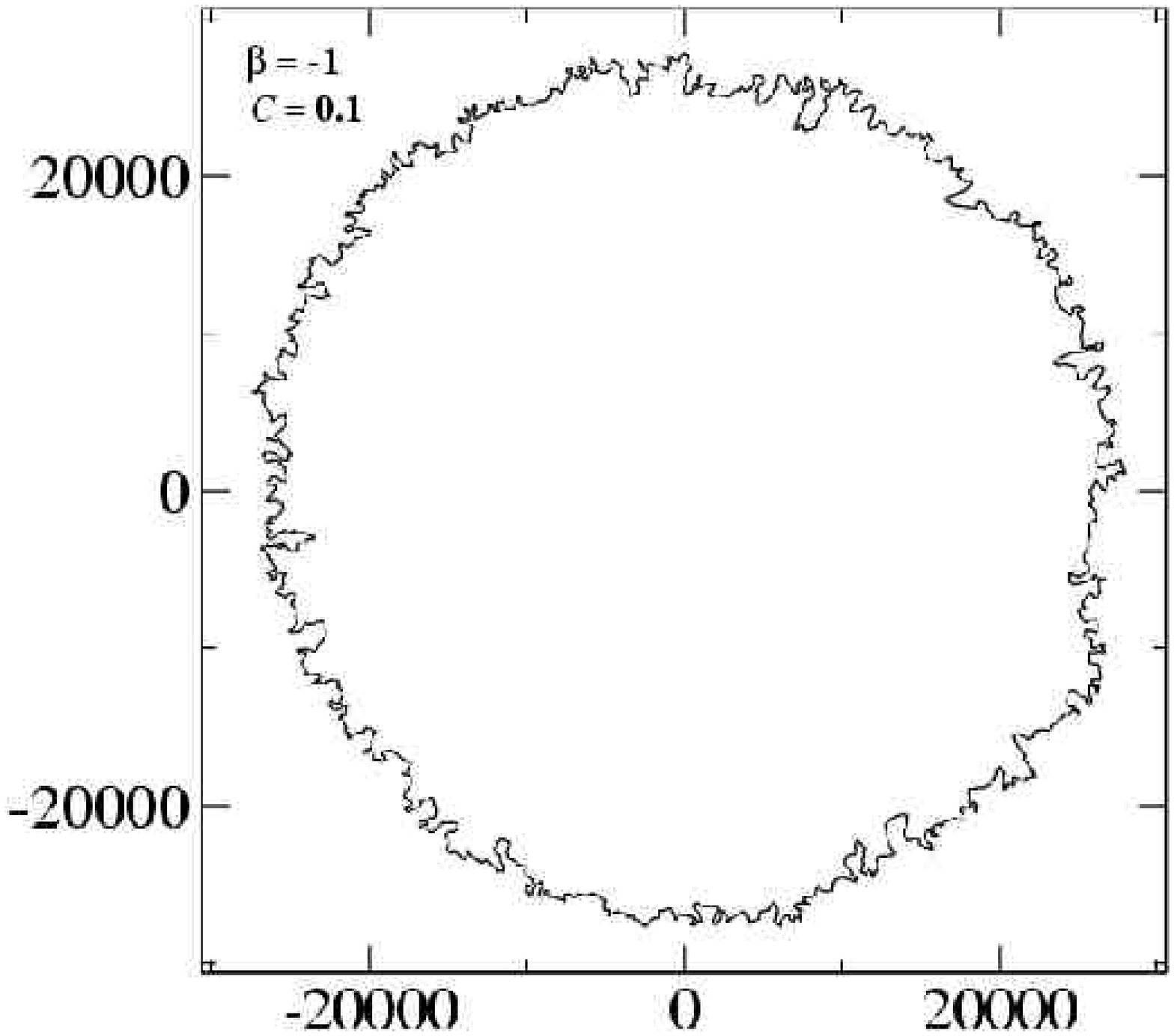}
\includegraphics[width=.33\textwidth]{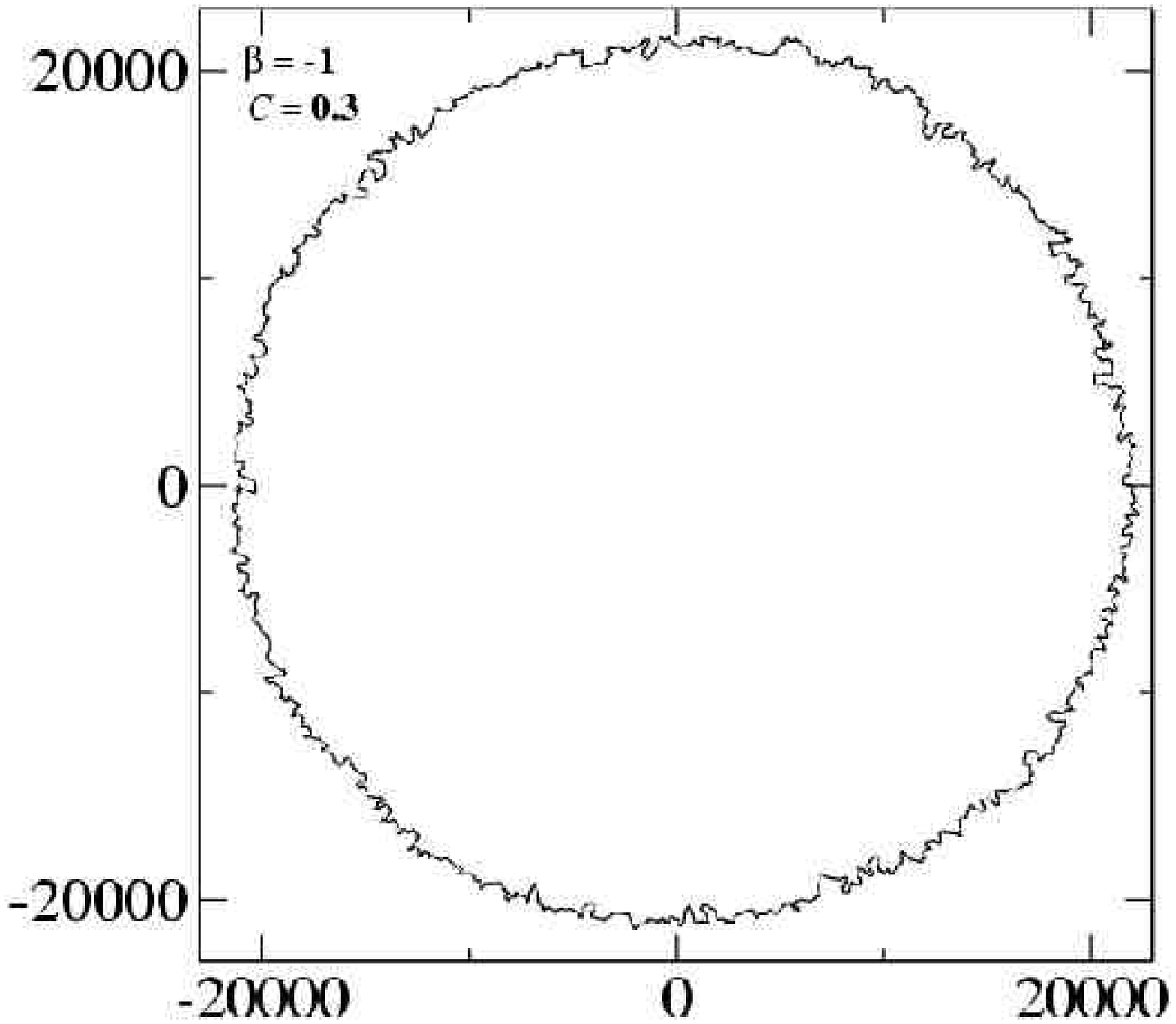}
\includegraphics[width=.33\textwidth]{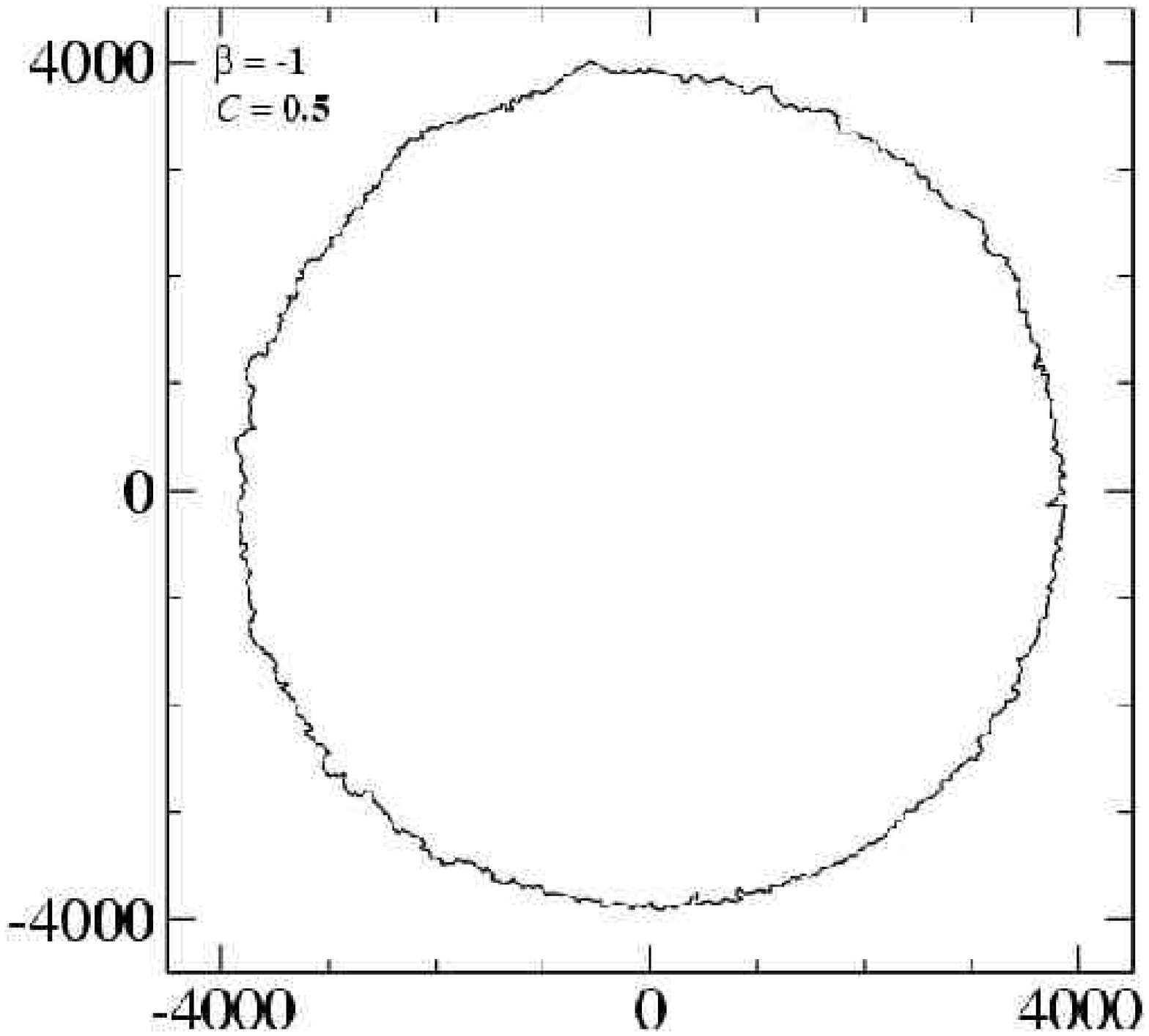}
\caption{Growth patterns for $\beta=-1$. Panels a-d: ${\cal C}=0.01,
0.1,0.3$ and 0.5 respectively.}
\label{clusters-1}
\end{figure} 
Therefore our aim is to find a line in the $\beta-\C.C$ phase diagram that 
separates fractal $D<2$ from 2-dimensional growth patterns. 
That such a line must exist we can convince ourselves by examining
the family of growth models that are seen for $\beta=-1$,
see Fig. \ref{clusters-1}. Obviously
these are 2-dimensional. The family of growth patterns obtained
for $\beta=0$ were shown in Fig. \ref{clusters0}, and as we
said above, there must be a cross over 2-dimensional patterns
in this family. Going up to 
$\beta=1$ we show the growth patterns in Fig. \ref{clustersbeta1}.
In this case the images indicate that for the lower values of
$\C.C$ the growth patterns are fractal, whereas for higher values
of $\C.C$ they become 2-dimensional. Thus the line of separation
that we seek in the $\beta-\C.C$ phase diagram appears to
cut the $\beta=1$ line. Finally, in Fig. \ref{clusters2} we present
the family of
growth patterns obtained for $\beta=2$.
It appears that the transition line intersects also the
$\beta=2$ line.

All the patterns exhibited in Figs. \ref{clusters0}, \ref{clusters-1}-\ref{clusters2}
are grown with a fixed size $\lambda_0$. Consequently, for $\beta>0$ the
actual mean size of the bumps in the physical space decreases as
the cluster grows, while it increases for $\beta<0$. This may
lead to worries, i.e. that for $\beta>0$ the growth arrests and that
for $\beta<0$ the increase in the size of the bumps leads to 
coverage of fjords, such that the 2-dimensional patterns shown
in Fig. \ref{clusters-1} would be an artifact. To disperse these
worries we have considered alternative growth algorithms with
varying the size of $\lambda_0$. The first such algorithm is obtained
by requiring that the total area covered in each layer of growth
is constant, i.e.
\begin{eqnarray}
&&\sum_{k=1}^p \lambda_{n+k} |{\Phi^{'(n)}} (e^{i\theta_{n+k}})|^2\nonumber\\=
&&\lambda_0(n)\sum_{k=1}^p |{\Phi^{'(n)}} (e^{i\theta_{n+k}})|^{-\beta}={\rm C} \ .
\label{rule1}
\end{eqnarray}
Note that for constant coverage $\C.C$ this rule coincides with fixed values of $\lambda_0$ for
$\beta=2$ (cf. Eq.(\ref{defC})).
In the second algorithm we choose the maximal size of the bump
in the physical plane to be constant from layer to layer:
\begin{equation}
\lambda_0(n)~\max_{k=1}^p \{|{\Phi^{'(n)}} (e^{i\theta_{n+k}})|^{-\beta}\}={\rm C} \ .
\label{rule2}
\end{equation}
This rule coincides with fixed values of $\lambda_0$ for $\beta=0$. We found
that in all cases the patterns shown above remain invariant to the
change of the algorithms. Thus we submit that the figures shown
can be fully trusted.

\begin{figure}
\centering
\includegraphics[width=.33\textwidth]{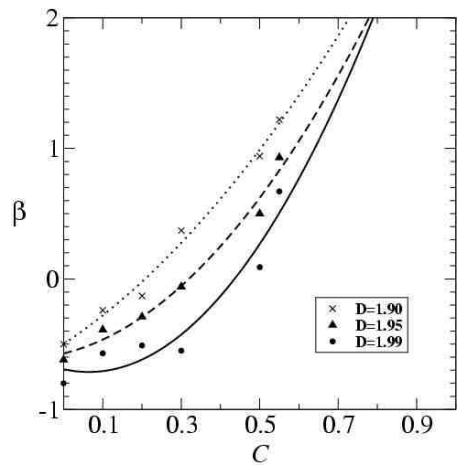}
\caption{The phase diagram in the $\beta-\C.C$ plane. The data
points in crosses, triangles and circles represent values of $\beta$ and $\C.C$ for
which the radius-area relationship predicts $D=1.90$, $D=1.95$ and $D=1.99$
respectively. The 
lines are quadratic fits. We propose that the region below the lines
represents 2-dimensional growth patterns, and see text for details.}
\label{phasedi}
\end{figure}

To find the line that separates fractal from 2-dimensional 
patterns we estimate the dimensions directly from log-log plots of
$F^{(n)}$ vs. $S$. We have seen above that such estimates are
{\em lower} bounds to the actual asymptotic dimension. As 
these logarithmic plots are invariably concave, we can
use the slope at the largest values of area available as
a measure for the lower bound on the dimension. In Fig. \ref{phasedi}
we show the three lines obtained by searching, for a given
value of $\C.C$, the value of $\beta$ for which for the first
time the dimension estimated from $F^{(n)}$ vs. $S$ crosses
the value $D=1.90$ (upper curve),  $D=1.95$ (middle line) and 
$D=1.99$ (lower curve). We propose that the last two lines
may very well be already beyond the true line that 
separates fractal from $D=2$ asymptotic dimension. 
From the discussion of Sect. IV A we cannot even exclude
the possibility that
the transition line obfuscates the $\beta=0$ line. All
the region below the lower line is almost surely representing
patterns of $D=2$, but we strongly believe that this is the
case also for the middle line. The lines were obtained 
by finding, as explained, the values of $\beta$ yielding
D=1.90, 1.95 and 1.99 respectively, and then fitting to the 
points a quadratic function. Next we extrapolated the
three fits to values of $\C.C$ that are not readily available
in our algorithm. The three fit lines intersect the $\beta=2$ 
line at $\C.C=0.73$, 0.78 and $\C.C=0.79$ respectively. We thus
propose that the value $\C.C=1$ for $\beta=2$ is comfortably
within the region of 2-dimensional patterns in this
phase diagram. 
\section{Conclusions}
We have presented a careful numerical study of a 2-parameter model of growth patterns
that generalizes and interpolates between Diffusion Limited
Aggregation and Laplacian Growth Patterns. The model gives rise
to a rich plethora of growth patterns, with fractal dimensions that
depend on the values of the parameters $\beta$ and $\C.C$. For
$\beta=0$ and $\C.C=0$ we obtain DLA. Laplacian Growth patterns
have $\beta=2$ and $\C.C=1$, but we cannot probe the value $\C.C=1$
within our algorithm. Since our aim, in part, is to demonstrate
that Laplacian Growth patterns are not fractal, we resorted to examining
the phase diagram $\beta-\C.C$. We established, on the basis of
scaling arguments, simulations and visual observations, that this
phase diagram contains a line of transition between fractal
and 2-dimensional growth patterns. We have estimated the position
of this line, and demonstrated that Laplacian
Growth patterns belong safely in the region of 2-dimensional
growth patterns.

One should point out that the statement that Laplacian Growth
are 2-dimensional does not mean that it is a growing disk. To the
eye the patterns can look fractal, and in fact radius-area 
log-log plots might initially even indicate that the dimension
is low, and maybe of the order of the dimension of DLA. Deep fjords
may exist in the structure. The relevant question is whether
the growing branches of the structure contain substance (area)
and whether this area is growing relatively with the growth
of the pattern. The growth pattern shown in Fig. \ref{clustersbeta1}
panel d is a case in point. It looks fractal to the naked eye,
but careful examination shows that the branches have area. Thus
one needs to decide whether this area is due to some ultraviolet
cutoff length, or does it grow systematically beyond what
is expected on the basis of the existence of such a cutoff.

Before closing we reiterate that our demonstration that Laplacian
Growth patterns are 2-dimensional is not direct. We cannot,
within our algorithm, grow $\C.C=1$ patterns. We therefore
leave this at the moment as a conjecture. It remains a theoretical 
challenge to show that this conjecture
is indeed provable by direct mathematical analysis. We also
leave for future work the question whether the $\beta=0$ line
represents 2-dimensional growth patterns for all $\C.C>0$.
Finally we propose that future work may make use of the 
fractal patterns along the line
$\C.C=0$, $\beta>0$ for further fundamental studies of DLA and
related phenomena.
 \acknowledgments
We thank Benny Davidovitch for a critical reading of the
manuscript, and for a number of useful comments.
This work has been supported in part by the
   European Commission under the TMR program, The Petroleum Research Fund 
and the Naftali and Anna
   Backenroth-Bronicki Fund for Research in Chaos and Complexity.
AL thanks the Minerva Foundation, Munich, Germany for financial support.
   
\end{document}